\documentclass[%
 aip,
 amsmath,amssymb,
 reprint,%
]{revtex4-1}

\usepackage{graphicx}% Include figure files
\usepackage{dcolumn}% Align table columns on decimal point
\usepackage{bm}% bold math
\usepackage{xr}
\makeatletter
\newcommand*{\addFileDependency}[1]{% argument=file name and extension
  \typeout{(#1)}
  \@addtofilelist{#1}
  \IfFileExists{#1}{}{\typeout{No file #1.}}
}
\makeatother

\usepackage[utf8]{inputenc}
\usepackage[T1]{fontenc}
\usepackage{mathptmx}
\usepackage{color}

\usepackage[page]{appendix}

\usepackage{caption}
\usepackage{subcaption}

\begin{document}

\preprint{AIP/123-QED}

\title[]{
A Self Consistent Field Formulation of Excited State Mean Field Theory
}
% Force line breaks with \\

\author{Tarini S. Hardikar}
\affiliation{
Department of Chemistry, University of California, Berkeley, California 94720, USA 
}
\author{Eric Neuscamman}
 \email{eneuscamman@berkeley.edu.}
\affiliation{
Department of Chemistry, University of California, Berkeley, California 94720, USA 
}
\affiliation{Chemical Sciences Division, Lawrence Berkeley National Laboratory, Berkeley, CA, 94720, USA}

\date{\today}% It is always \today, today,

\begin{abstract}
We show that, as in Hartree Fock theory, the orbitals for
excited state mean field theory can be optimized via a
self-consistent one-electron equation in which
electron-electron repulsion is accounted for through
mean field operators.
In addition to showing that this excited state ansatz
is sufficiently close to a mean field product state
to admit a one-electron formulation, this approach
brings the orbital optimization speed to within
roughly
a factor of two of ground state mean field theory.
The approach parallels Hartree Fock theory in multiple
ways, including the presence of
a commutator condition,
a one-electron mean-field working equation,
and acceleration via
direct inversion in the iterative subspace.
When combined with a configuration interaction
singles Davidson solver for the excitation coefficients,
the self consistent field formulation
dramatically reduces the cost of the theory compared
to previous approaches based on quasi-Newton descent.
\end{abstract}

\maketitle

\section{Introduction}
\label{sec:intro}

Hartree Fock (HF) theory \cite{Szabo-Ostland,MolElecStruc}
is so immensely useful in large part due to the rigorous
and convenient link it provides between a qualitatively
correct many-electron description and an affordable
and more
intuitive one-electron equation.
The link it makes is rigorous in that, when solved, its one-electron
equation guarantees that the many-electron description underneath it
is optimal in a variational sense,
meaning that the energy is made stationary with respect to
changes in the wave function.
The link is also convenient, because many-electron properties
like the energy
can be evaluated in terms of inexpensive one-electron quantities,
and because solving a one-electron equation, even one with
mean field operators that must be brought to self-consistency,
is in most cases easier and less expensive than
a direct minimization of the many-electron energy.
The fact that this useful link is possible at all owes much to the
simplicity of the Slater determinant many-electron wave function
on which HF theory is built.
Essentially, the Slater determinant is as close as we can get to a
truly mean field, correlation-free Hartree product ansatz
while still capturing the important effects of Pauli correlation.
Happily, this single step away from a product state does not prevent
a useful and intuitive formulation in terms of a self-consistent
one-electron equation in which mean field operators account for
electron-electron coulomb repulsion.

In this paper, we will show how excited state mean field (ESMF) theory
\cite{Shea2018}
can also be formulated in terms of a one-electron mean field
equation that, when solved self consistently, produces optimal orbitals.
As in HF theory, this formulation is possible thanks to the ansatz hewing
closely to the mean field limit:  ESMF takes only one additional step away
from a truly mean field product state by adding the open-shell correlation
that arises in an excitation on top of the Pauli correlations already
present in the ground state.
Perhaps most importantly, the resulting one-electron equation that
determines the optimal orbitals can, like the Roothaan equations,
be solved by iteratively updating a set of mean field operators
until they are self-consistent with the orbital shapes.
As we will see, when accelerated by direct inversion in the
iterative subspace (DIIS), \cite{pulay1982diis}
this self consistent field (SCF) approach brings the orbital
optimization cost down to within
a factor of two of HF theory, and significantly lowers
the overall cost of ESMF theory compared to previous approaches.
Given that ESMF offers a powerful platform upon which
to construct excited-state-specific correlation theories
\cite{Zhao2019dft, Shea2020gvp, Clune2020topesmp2}
and that it has recently been shown to out-compete other low-cost methods
like configuration interaction singles (CIS) and density functional theory
in the prediction of charge density changes, \cite{zhao2020esmf}
this acceleration of the theory and simplification of its
implementation should prove broadly useful.

While recent work has provided an improved ability to
optimize the ESMF ansatz via the nonlinear minimization of a generalized
variational principle (GVP), \cite{Shea2020gvp, zhao2020esmf}
the current lack of an SCF formulation stands in
sharp contrast to the general state of affairs
for methods based on Slater determinants.
Even in contexts outside of standard HF for ground states,
SCF procedures are the norm
rather than the exception when it comes to optimizing
Slater determinants' orbitals.
Indeed, among many others, the
$\Delta$SCF, \cite{bagus1965scf,Pitzer1976scf,argen1991xray,Gill2009dscf}
restricted open-shell Kohn Sham, \cite{Shaik1999,Kowalczyk2013}
constrained density functional theory, \cite{VanVoorhis2005cdft}
ensemble density functional theory,
\cite{theophilou1979,kohn1986quailocal,gross1988density}
projected HF, \cite{jimenez2012PHF}
and $\sigma$-SCF \cite{vanVoorhis2017sigmaSCF,Voorhis2019}
methods all favor SCF optimization approaches.
Although the direct minimization of a GVP or the norm of the energy
gradient \cite{hait2020oo} offers protection against a Slater determinant's
``variational collapse'' to the ground state or lower excited states,
this rigorous safety comes at some cost to efficiency.
It is not for nothing that direct energy minimization methods, although
available, \cite{VanVoorhis2002gdm} are not the default HF optimization
methods in quantum chemistry codes.
In cases where they prove stable, SCF approaches are typically more efficient.
In the case of the ESMF anstatz, an SCF approach is also at risk
of collapse to an undesired state, but,
even in such troublesome cases, a brief relaxation of the orbitals
by SCF may still offer a low-cost head start for the
direct minimization of a GVP.
In cases where an SCF approach to ESMF is stable, history strongly
suggests that it will be more efficient than nonlinear minimization.
In short, our preliminary data agree with history's suggestion.

\section{Theory}
\label{sec:theory}

\subsection{Hartree-Fock Theory}
\label{sec::hf}

To understand how an SCF formulation of ESMF
theory comes about, it is useful to first review the formulation of
HF theory and in particular how its condition for optimal orbitals
can be written as a commutator between a mean field operator and
a one-body reduced density matrix (RDM).
In HF theory, the energy of the Slater determinant
$\Psi_{SD}$ is made stationary with respect
to changes in the orbital variables, which is the Slater determinant's
approximation of the more general condition that an exact energy
eigenstate will have an energy that is stationary with respect to
any infinitesimal variation in the wave function.
For convenience, and without loss of generality, the molecular orbitals
are constrained by Lagrange multipliers to be orthonormal.
\cite{Szabo-Ostland}
For Restricted Hartree Fock (RHF),
the resulting Lagrangian
\begin{align}
\label{eqn:rhf_lagrangian}
%L_{RHF} &=       \left \langle\Psi_{SD}|H|\Psi_{SD}\right \rangle 
%         + 2 \hspace{0.5mm} \mathrm{tr}
%         \left[ (\bm{I} - \bm{C}^T \bm{S} \bm{C}) \bm{\epsilon} \right]
%\notag \\
L_{RHF}  &=       E_{RHF} 
         + 2 \hspace{0.5mm} \mathrm{tr}
         \left[ (\bm{I} - \bm{C}^T \bm{S} \bm{C}) \bm{\epsilon} \right]
\end{align}
in which $\bm{C}$ is the matrix whose columns hold the molecular orbital
coefficients, $\bm{S}$ is the atomic orbital overlap matrix,
$\bm{I}$ is the identity matrix,
$\bm{\epsilon}$ is the symmetric matrix of Lagrange multipliers,
$\mathrm{tr}[]$ is the matrix trace operation,
and $E_{RHF}$ is the RHF energy (given below),
is then made stationary by setting derivatives with respect to $\bm{C}$
equal to zero.
After some rearrangement, \cite{Szabo-Ostland}
this condition can be formulated into the famous Roothaan equations,
\begin{align}
\label{eqn:rhf_roothaan}
\big( \bm{h} + \bm{W}\left[\bm{A}\right] \big) \bm{C}
  = \bm{S} \bm{C} \bm{\epsilon}
\end{align}
in which $\bm{h}$ is the matrix representation of the one-electron
components of the Hamiltonian in the atomic orbital basis
and $\bm{W}$ is interpreted as a mean field approximation for
electron-electron repulsions.
Of course, this mean field repulsion depends on the orbital shapes,
causing the operator $\bm{W}$ to be a function of $\bm{A}$, the
Aufbau determinant's 1-body $\alpha$-spin RDM.
In what comes below we will consider RDMs and other matrices
in both the atomic orbital (AO) and molecular orbital (MO) bases,
and will adopt the notation that a matrix with no superscript
(e.g.\ $\bm{A}$) refers to the AO representation, while the MO
representation is explicitly denoted as such (e.g.\ $\bm{A}^{(MO)}$).
The closed-shell Aufbau determinant's RDM has the form
\begin{align}
\label{eqn:rhf_rdm}
\bm{A}^{(MO)} = \bm{I}_o
\qquad
\bm{A} = \bm{C} \bm{A}^{(MO)} \bm{C}^T
\end{align}
where the matrix $\bm{I}_o$ has ones on
the first $n_o$ elements of its diagonal and zeros elsewhere
($n_o$ is the number of occupied molecular orbitals).
Although in many contexts it is useful to separate
the restricted HF (RHF) mean field electron-electron
repulsion operator $\bm{W}[\bm{A}]=2\bm{J}[\bm{A}]-\bm{K}[\bm{A}]$
into its ``coulomb'' $\bm{J}$ and ``exchange'' $\bm{K}$ components, 
\begin{align}
\label{eqn:J}
J[\mathcal{\gamma}]_{pq}&=\sum_{rs}{\mathcal{\gamma}_{rs}(rs|pq)} \\
\label{eqn:K}
K[\mathcal{\gamma}]_{pq}&=\sum_{rs}{\mathcal{\gamma}_{rs}(pr|qs)},
\end{align}
defined here using the two-electron integrals in 1122 order,
this separation is not necessary at present and so we will
work instead in terms of the combined mean field operator $\bm{W}$.

Now, while the Roothaan equation has both an intuitive appeal
as a one-electron Schr\"{o}dinger equation and a practical appeal
as a convenient setup for an SCF cycle based on the
efficient numerical diagonalization of
a symmetric generalized eigenvalue problem, it is not the only
way to formulate HF theory's central requirement of Lagrangian
stationarity.
Noting that only the first $n_o$ columns of $\bm{C}$ affect the
ansatz, we can right-multiply Eq.\ (\ref{eqn:rhf_roothaan})
by $\bm{I}_o=\bm{A}^{(MO)}$ to focus our attention
on them while at the same time left-multiplying by $\bm{C}^T$
to eliminate the overlap matrix, which results in
\begin{align}
\label{eqn:rhf_modified_roothan}
\bm{F}^{(MO)} \bm{A}^{(MO)} =
\bm{C}^T \big( \bm{h} + \bm{W} \big) \bm{C} \bm{A}^{(MO)}
  = \bm{\epsilon} \bm{I}_o
\end{align}
where we have made the usual definition of the Fock operator.
\begin{align}
\label{eqn:rhf_fock_op}
\bm{F}^{(MO)} = \bm{C}^T\bm{F}\bm{C}
= \bm{C}^T \big( \bm{h} + \bm{W} \big) \bm{C}
\end{align}
If we ensure that we work in the canonical representation,
\cite{Szabo-Ostland}
the matrix $\bm{\epsilon}$ will be diagonal, and so
Eq.\ (\ref{eqn:rhf_modified_roothan})
essentially says that the product
$\bm{F}^{(MO)} \hspace{0.3mm} \bm{A}^{(MO)}$
must produce a symmetric matrix.
We may enforce this requirement by setting the difference
between this product and its transpose equal to zero,
which leads to a commutator condition for Lagrangian
stationarity that can be used as an alternative to the
Roothaan equation
when optimizing orbitals.
\cite{McWeeny, pulay1982diis} 
\begin{align}
\label{eqn:rhf_commutator}
\big[ \hspace{0.6mm} \bm{C}^T\bm{F}\bm{C},
      \hspace{0.6mm} \bm{A}^{(MO)} \hspace{0.6mm} \big] = 0
\end{align}
If we consider the HF energy expression
\begin{align}
\label{eqn:rhf_energy}
E_{RHF} = \mathrm{tr}\big[ (2\bm{h} + \bm{W}) \bm{A} \big]
\end{align}
alongside the Fock operator definition $\bm{F}=\bm{h}+\bm{W}$,
we see a nice connection between the commutator condition and the
energy.
Specifically, if one halves the one-electron component of the
mean field operator whose trace with the density yields
the energy, the resulting operator ($\bm{F}$ in this case)
must, when put in the MO basis, commute with the MO basis representation
of the density matrix in order for the Lagrangian to be stationary.
With this connection pointed out, we now turn our attention to ESMF theory,
where a generalization of Eq.\ (\ref{eqn:rhf_commutator}) yields a
useful SCF formulation for orbital optimization.

\subsection{Excited State Mean Field Theory}
\label{sec::esmf}

Like HF theory, the energy expression for the ESMF ansatz for
a singlet excited state can be written in terms of traces between
mean field operators and density-like matrices.
In particular, if we take the simple version of the singlet ESMF ansatz
in which the Aufbau coefficient is set to zero,
\begin{align}
\label{eqn:esmf_ansatz}
\left. |\Psi_{ESMF}\right\rangle
= \sum_{ia}
  t_{ia} \left|\hspace{0.3mm}{}^{a_{\uparrow}}_{i_{\uparrow}}\right\rangle
+ t_{ia} \left|\hspace{0.3mm}{}^{a_{\downarrow}}_{i_{\downarrow}}\right\rangle,
\end{align}
where $\bm{t}$ is the matrix of CIS-like configuration interaction
coefficients and
$\left|\hspace{0.3mm}{}^{a\uparrow}_{i\uparrow}\right\rangle$ is the
Slater determinant resulting from an
$\hspace{1mm}i\rightarrow a\hspace{1mm}$
$\alpha$-spin excitation
out of the Aufbau determinant (note we do not say the HF determinant,
as we are not in the HF MO basis),
then the ESMF singlet energy amounts to four traces between mean field
operators and density-like matrices.
\begin{align}
\notag
E_{ESMF} =
\hphantom{+} \hspace{0.5mm} \mathrm{tr}\big[&
                 \hspace{0.7mm} ( \hspace{0.5mm} 2\bm{h} \hspace{0.3mm}
                                 + \bm{W}[\bm{A}] \hspace{0.7mm} )
                        \hspace{1mm} \mathcal{\gamma}
         \hspace{1mm} \big]
%\notag
 + \hspace{0.5mm} \mathrm{tr}\big[ \hspace{0.5mm} \bm{W}[\bm{D}]
                          \hspace{0.5mm} \bm{A} \hspace{0.5mm} \big] \\
%\notag
\label{eqn:esmf_energy}
 + \hspace{0.5mm} \mathrm{tr}\big[& \hspace{0.5mm} \bm{W}[\bm{T}]
                          \hspace{0.5mm} \bm{T}^T \hspace{0.5mm} \big]
 + \hspace{0.5mm} \mathrm{tr}\big[ \hspace{0.5mm} (\bm{W}[\bm{T}])^T
                          \hspace{0.5mm} \bm{T} \hspace{0.5mm} \big]
\end{align}
Here $\mathcal{\gamma}$ is the one-body alpha-spin
RDM for the ESMF ansatz.
\begin{align}
\label{eqn:esmf_rdm}
\gamma^{(MO)}
= \bm{I}_o + \left(\begin{array}{c|c}
   -\bm{t}\hspace{0.4mm}\bm{t}^T & 0 \\
     \hline
  0  & \bm{t}^T\bm{t}\rule{0pt}{3.8mm}
\end{array}\right)
\qquad
\gamma = \bm{C} \gamma^{(MO)} \bm{C}^T
\end{align}
The matrix $\bm{A}$ is the Aufbau determinant's one-body RDM,
as in Eq.\ (\ref{eqn:rhf_rdm}).
The difference between these density matrices we define as
$\bm{D}=\gamma-\bm{A}$.
Finally, $\bm{T}$ is the non-symmetric matrix
that, in its MO representation, has the $\alpha$-spin
transition density matrix
between the Aufbau determinant and the ESMF ansatz
(which is as for CIS just $\bm{t}\hspace{0.2mm}$)
in its upper-right corner.
\begin{align}
\label{eqn:esmf_tdm}
\bm{T}^{(MO)}
= \left(\begin{array}{c|c}
   0 & \bm{t} \\
     \hline
  0  & 0 \rule{0pt}{3.8mm}
\end{array}\right)
\qquad
\bm{T}=\bm{C}\bm{T}^{(MO)}\bm{C}^T
\end{align}

With the ESMF energy written in terms of one-body mean field
operators and density-like matrices, we can now present our
central result, in which the stationarity conditions for
the ESMF Lagrangian
\begin{align}
\label{eqn:esmf_lagrangian}
L_{ESMF} = E_{ESMF}
         + 2 \hspace{0.5mm} \mathrm{tr}
         \left[ (\bm{I} - \bm{C}^T \bm{S} \bm{C}) \bm{\epsilon} \right]
\end{align}
with respect to orbital variations are written in a one-electron
equation that admits an SCF-style solution.
We begin, as in HF theory, by setting the (somewhat messy) derivatives
$\partial L_{ESMF} / \partial \bm{C}$ equal to zero.
With some care, this condition can be organized into
\begin{align}
\notag
&\big( \bm{h}+\bm{W}[\bm{A}] \big) C \gamma^{(MO)}
+ \bm{W}[\bm{D}] C A^{(MO)} \\
&\hspace{6mm}
+ \bm{W}[\bm{T}] C (T^{(MO)})^T
+ (\bm{W}[\bm{T}])^T C T^{(MO)}
= \bm{S} \bm{C} \bm{\epsilon}
\label{eqn:esmf_roothaan}
\end{align}
whose structure is similar to but also notably different
from the analogous HF expression in Eq.\ (\ref{eqn:rhf_roothaan}).
The formal difference is that there are now four terms on the
left hand side, one for each trace in the energy expression.
The practical difference is that the ESMF equation is not
an eigenvalue problem, and it is not obvious that it can be
reorganized into one due to the incompatible kernels of the
matrices $\gamma^{(MO)}$, $A^{(MO)}$, and $T^{(MO)}$.
Thus, it is at present not clear whether this ESMF equation
can offer the same spectral information that the Roothaan
equation provides for HF.
Nonetheless, for orbital optimization,
we have found a convenient alternative
by transforming this stationary condition into commutator
form by following the same steps that took us from
Eq.\ (\ref{eqn:rhf_roothaan}) to Eq.\ (\ref{eqn:rhf_commutator})
in HF theory.
Defining $\bm{F}_A=\bm{h}+\bm{W}[\bm{A}]$,
the result is that the Lagrangian stationary condition can be
written as
\begin{align}
\notag
0 = & \hspace{0.3mm} \big[ \hspace{1mm} \bm{C}^T\bm{F}_A\bm{C},
      \hspace{1mm} \mathcal{\gamma}^{(MO)} \hspace{1mm} \big] \\
\notag
& + \big[ \hspace{1mm} \bm{C}^T\bm{W}[\bm{D}]\bm{C},
      \hspace{1mm} \bm{A}^{(MO)} \hspace{1mm} \big] \\
\notag
& + \big[ \hspace{1mm} \bm{C}^T\bm{W}[\bm{T}]\bm{C},
      \hspace{1mm} (\bm{T}^{(MO)})^T \hspace{1mm} \big] \\
& + \big[ \hspace{1mm} \bm{C}^T(\bm{W}[\bm{T}])^T\bm{C},
      \hspace{1mm} \bm{T}^{(MO)} \hspace{1mm} \big].
\label{eqn:esmf_commutator}
\end{align}
It is interesting that the same pattern holds as in the HF
case:  the commutator condition has one commutator per
trace in the energy expression, and the mean field operators
(with any one-electron parts halved) are again paired with
the same density-like matrices as in the energy traces.
We find this pattern especially interesting in light of the
fact that it does not simply follow that each trace produces
one commutator.
Instead, cancellations of terms coming from derivatives on
different traces are needed to arrive at the commutators
above, and so we do wonder whether this is a happy accident
or whether there is an underlying reason to expect such
cancellations.

\subsection{Self Consistent Solution}
\label{sec::scs}

Either way, Eq.\ (\ref{eqn:esmf_commutator}) forms the
basis for an efficient SCF optimization of the ESMF
orbitals.
Assuming that we are a small orbital rotation 
away from stationarity, we insert the rotation
$\bm{C}\rightarrow\bm{C}\mathrm{exp}(\bm{X})$ into our
commutator condition and then expand the exponential
and drop all terms higher than linear order in the
anti-symmetric matrix $\bm{X}$.
The result is a linear equation for $\bm{X}$
(see Eq.\ (\ref{eqn:linear_eq_x}) in the
Supplementary Material)
which we solve via
the iterative GMRES method.
Note that, if desired, one can control the maximum
step size in $\bm{X}$ by simply stopping the GMRES
iterations early if the norm of $\bm{X}$ grows beyond
a user-supplied threshold.
This may be desirable,
as we did after all
assume that only a small rotation was needed and our
linearization of the equation prevents us from trusting
any proposed rotation that is large in magnitude.
In parallel to SCF HF theory, which holds $\bm{F}$ fixed
while solving the Roothaan equation for new orbitals,
we hold $\bm{F}_A$, $\bm{W}[\bm{D}]$, and $\bm{W}[\bm{T}]$
fixed while solving our linear equation.
Thus, although the modified GMRES solver is not
as efficient as the dense eigenvalue solvers used for HF
theory, it remains relatively inexpensive as it does
not does not involve any Fock builds and so does not have
to access the two-electron integrals.
(\begin{small}\textit{Technical note:
in practice, we can speed up the GMRES solver
considerably by preconditioning it with a diagonal
approximation to the linear
transformation that is set to one for $\bm{X}$ elements
in the occupied-occupied and virtual-virtual
blocks (since these are expected to play little role
in the orbital relaxation) and, in the other blocks,
replaces $\bm{C}^T\bm{F}_A\bm{C}$ with its diagonal,
replaces $\gamma^{(MO)}$ with $\bm{I}_o$,
and neglects $\bm{W}[\bm{D}]$ and $\bm{W}[\bm{T}]$
(see Supplementary Material for the explicit form).
DIIS is also effective when
we take Eq.\ (\ref{eqn:esmf_commutator})
transformed into the AO basis as the error vector
and the $\bm{F}_A$, $\bm{W}[\bm{D}]$ and $\bm{W}[\bm{T}]$
matrices as the DIIS parameters.
We use both of these accelerations in all calculations.}\end{small})
Only after the linear equation is solved and the orbitals
are updated do we rebuild the three mean field operators,
and so each overall SCF iteration requires just three
Fock builds, which, as they can be done during the same
loop over the two-electron integrals, come at a cost
that is not much different than HF theory's single
Fock build.
This arrangement contrasts sharply with the nine Fock
builds and two integral loops that are necessary to
form the analytic derivative of the energy with respect
to $\bm{C}$ that is used in descent-based orbital
optimization. \cite{zhao2020esmf}
In summary, the ESMF orbitals, like the HF orbitals,
can be optimized particularly efficiently
via the self-consistent solution of a one-electron
mean field equation.

Although this exciting result makes clear that the ESMF
ansatz really does hew closely enough to the mean field
product-state limit for one-electron mathematics to
be of use, there are a number of questions we should
now address.
First, and we will go into more detail on this point
in the next paragraph,
is the SCF approach actually faster than descent?
The answer, at least in simple systems, is a resounding yes.
Second, what of the configuration interaction coefficients
$\bm{t}$?
At present, we optimize them in a two-step approach, in which
we go back and forth between orbital SCF solutions
and CIS calculations (taking care to include the new terms
that arise for CIS when not in the HF MO basis)
until the energy stops changing.
In future, more sophisticated approaches that provide
approximate coupling between these optimizations may be
possible, as has long been true in multi-reference
theory. \cite{kreplin2020mcscf}
Third, what physical
roles can we ascribe to the different mean field operators
that appear in the SCF approach to ESMF?
The operator $\bm{F}_A$ obviously carries the lion's share
of the electron-electron repulsion, as it is the only
mean field operator derived from a many-electron density
matrix.
Indeed, $\bm{W}[\bm{D}]$ and $\bm{W}[\bm{T}]$ represent
repulsion from one-electron densities, and so they
cannot provide the bulk of the electron-electron repulsion.
Thus, we suggest that it is useful to view
$\bm{F}_A$ as a good starting point that includes
the various repulsions between electrons not involved
in the excitation but that gets the repulsions
affected by the excitation wrong.
$\bm{W}[\bm{D}]$ and $\bm{W}[\bm{T}]$ then act as
single-electron-density corrections to this starting
point.
If one considers the simple case in which we ignore
all electrons other than the pair involved in the
excitation (e.g.\ consider the HOMO/LUMO excitation
in H$_2$), then a close inspection reveals that
$\bm{W}[\bm{D}]$ eliminates the spurious HOMO-HOMO
repulsion that is present in the first
trace of the energy expression, while
the $\bm{W}[\bm{T}]$ terms bring the excited
electron pair's repulsion energy into alignment
with the actual repulsion energy that results
from the singlet's equal superposition of two
open-shell determinants.

\begin{table}[t]
\caption{\label{tab:h2o_homo_lumo}Convergence of 
   SCF- and GVP-based ESMF
   for the HOMO/LUMO excitation of cc-pVDZ H$_2$O.
   Initial values for $\bm{t}$ and $\bm{C}$ are set
   to the two-determinant HOMO/LUMO open shell
   singlet and the RHF orbitals, respectively.
   For SCF, the two-step method toggled between
   CIS and SCF calculations, with CIS going first.
   As the guess is quite
   good in this system, the GVP optimization
   set $\mu=0$ right away and so amounted
   to a BFGS minimization of the energy gradient norm.
   At various points during each optimization
   (measured both by the cumulative number of loops
    over the TEIs and by the wall time)
   we report the energy error $\Delta E$ compared
   to the fully converged energy.
   Both calculations used a single
   core on a 2015 MacBook Air.
}
\begin{tabular}{c c c c c c c}
\hline\hline
\multicolumn{3}{c}{SCF ESMF} &
$\quad$ &
\multicolumn{3}{c}{GVP ESMF \rule{0pt}{3.2mm}} \\
TEI Loops & Time (s) & $\Delta E$ (a.u.) &
 &
TEI Loops & Time (s) & $\Delta E$ (a.u.) \\
\hline
 10 & 0.007 & 0.062605 & &  76 & 0.397 & 0.003761 \rule{0pt}{3.2mm} \\
 20 & 0.025 & 0.000032 & & 150 & 0.783 & 0.000654 \\
 30 & 0.033 & 0.000004 & & 226 & 1.187 & 0.000184 \\
 40 & 0.054 & 0.000000 & & 300 & 1.579 & 0.000001 \\
\hline\hline
\vspace{0.1mm}
\end{tabular}
\end{table}

\begin{table}[t]
\caption{\label{tab:orb_timing}Total time
   in seconds
   and number of iterations $n_i$ taken for
   the orbital optimization in the ground state
   (for RHF) or the excited state (for SCF-based ESMF)
   to get within 5$\mu$E$_h$ of its fully converged value.
   The RHF and ESMF methods rely on the same underlying
   Fock build code, both use DIIS, and both used one
   core on a 2015 MacBook Air.
   For ESMF, only the orbitals are optimized, with
   $\bm{t}$ set to the HOMO/LUMO open-shell singlet
   and the initial guess for $\bm{C}$ set to the
   RHF orbitals.
   For RHF, the eigen-orbitals of the one-electron
   Hamiltonian were used as the initial guess for $\bm{C}$.
   Times do not include the generation of one-
   and two-electron AO integrals, which are the same
   for both methods.
}
\begin{tabular}{l c c c c c}
\hline\hline
Molecule \hspace{8mm} &
\hspace{1mm} Basis \hspace{1mm} &
\hspace{1mm} RHF (s) \hspace{0mm} &
\hspace{0mm} $n_i$ \hspace{1mm} &
\hspace{2mm} ESMF  (s) \hspace{0mm}\rule{0pt}{3.2mm} &
\hspace{0mm} $n_i$ \hspace{1mm} \\
\hline
water        & cc-pVTZ & 0.087 &  8 & 0.185 &  6\rule{0pt}{3.2mm} \\
formaldehyde & cc-pVTZ & 0.424 & 11 & 0.862 &  8  \\
ethylene     & cc-pVTZ & 0.903 &  8 & 1.735 &  6  \\
toluene      & cc-pVDZ & 4.366 & 19 & 6.835 & 11  \\
\hline\hline
\vspace{0.1mm}
\end{tabular}
\end{table}

\section{Results}
\label{sec:results}

\subsection{Efficiency Comparisons}
\label{sec::efficiency}

Returning now to the question of practical efficiency,
we report in Table \ref{tab:h2o_homo_lumo} the convergence of the energy
for the HOMO/LUMO excitation in the water molecule
for both SCF-based and GVP descent-based ESMF
(note all geometries can be found in the
Supplementary Material).
Whether one measures by the number of times the expensive
two-electron integral (TEI) access must be performed or by the wall
time, the two-step SCF approach is dramatically
more efficient than GVP-based descent in this case.
(The keen-eyed observer will notice that in the SCF case,
the TEI loop count and the wall time do not increase
at the same rate, which is due to the CIS iterations having
many fewer matrix operations to do as compared to SCF
in between each access of the TEIs.)
If we focus in on just the orbital optimization, as shown
in Table \ref{tab:orb_timing}, we find that the SCF
approach for ESMF is almost as efficient as ground state
HF theory.
In practice, of course, we also want to optimize $\bm{t}$,
and for now we rely on the two-step approach, as used
in Table \ref{tab:h2o_homo_lumo}.

%\subsection{Core Excitation}
%\label{sec::core}

While the SCF approach has clear advantages in simple
cases, the GVP is still expected to be essential
for cases in which the SCF approach may not be stable.
For example, without implementing an interior root solver
or freezing an open core
(and we have not done either), Davidson-based CIS would be
problematic for a core excitation.
However, as shown in Table \ref{tab:h2o_core},
a combination of an initial SCF optimization of the orbitals
followed by a full GVP optimization of $\bm{t}$ and $\bm{C}$
together is quite effective.
In this case, the SCF approach brings the energy
close to its final value, converging to an energy that
is too low by 54 $\mu$E$_h$ (remember, excited states
do not have any upper bound guarantee, even when a variational
principle like energy stationarity or the GVP is in use).
From this excellent starting point, the GVP's
combined optimization of $\bm{C}$ and $\bm{t}$ converges
quickly to the final energy, needing just ten
gradient evaluations to get within 1 $\mu$E$_h$.
In contrast, if the initial SCF orbital optimization
is omitted, the GVP coupled optimization
requires hundreds of gradient evaluations
(exactly how many depends on the choice
for $\omega$ and how $\mu$ is stepped down to zero)
\cite{Shea2020gvp}
to reach the same level of convergence, and was only able
to converge to the correct state at all by setting
$\mu$ to 0.5 and $\omega$ 0.08 E$_h$ lower than the final
energy for the initial iterations to avoid converging
to a higher-energy core excitation.
Especially interesting is the fact that,
if we move to the aug-cc-pVTZ basis, the
ESMF predictions for the two lowest
core excitations in H$_2$O are
534.3 and 536.2 eV, which are quite close to
the experimental values \cite{schirmer1993}
of 534.0 and 535.9 eV and which match
the delta between them even more closely.
Thus, even in cases where the SCF approach would be
difficult to use on its own, it can offer significant
benefits in partnership with direct minimization.

\begin{table}[t]
\caption{\label{tab:h2o_core}Convergence of
   the energy for the lowest singlet core excited
   state of H$_2$O in the aug-cc-pVDZ basis.
   Initial values for $\bm{t}$ and $\bm{C}$ are set
   to the two-determinant 1s$\rightarrow$LUMO
   open shell singlet and the RHF orbitals, respectively.
   An initial SCF optimization converged after
   10 iterations (involving one TEI loop each),
   after which GVP-based BFGS descent (again with $\mu$
   set immediately to zero) was started from
   the SCF result (the GVP requires 2 TEI loops per
   gradient evaluation).
   We report the energy error $\Delta E$ compared
   to the fully converged energy as a function
   of the cumulative wall time and the cumulative
   number of TEI loops.
   The calculation used a single
   core on a 2015 MacBook Air.
}
\begin{tabular}{c c r}
\hline\hline
TEI Loops & \hspace{2mm} Time (s) \hspace{2mm} &
$\Delta E$ (a.u.) \rule{0pt}{3.2mm} \\
\hline
\multicolumn{3}{l}{Start with SCF: \rule{0pt}{3.2mm}} \\
  5 & 0.163 &  0.008698 \hspace{0.3mm} \\
 10 & 0.267 & -0.000054 \hspace{0.3mm} \\
\multicolumn{3}{l}{Switch to GVP:} \\
 20 & 0.435 &  0.000002 \hspace{0.3mm} \\
 30 & 0.604 &  0.000001 \hspace{0.3mm} \\
\hline\hline
\vspace{0.1mm}
\end{tabular}
\end{table}

\begin{figure}[t]
    \centering
    \includegraphics[width=0.48\textwidth]{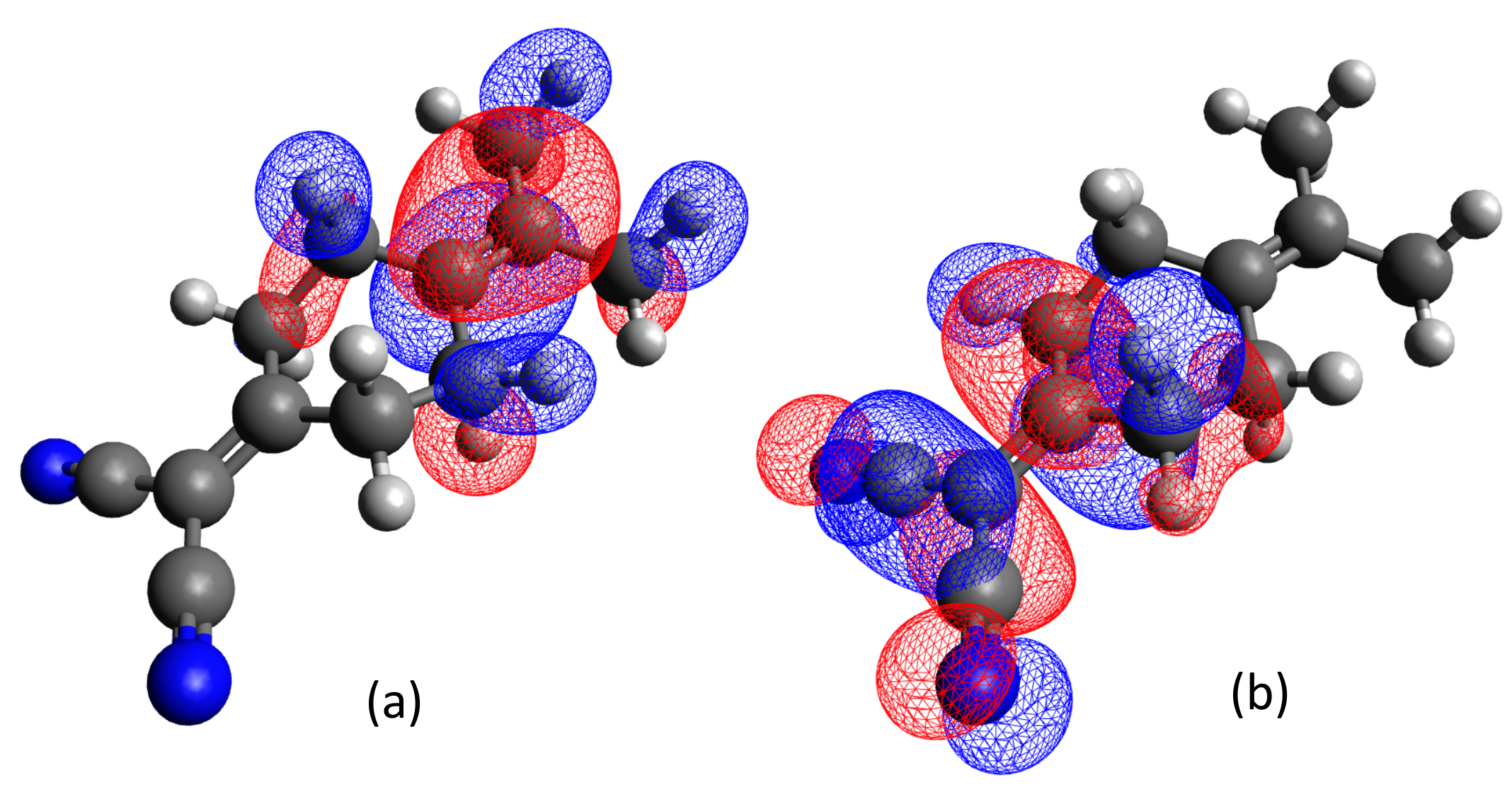}
    \caption{
    Donor (a) and acceptor (b) orbitals for the lowest
    charge transfer state in the PYCM molecule
    as predicted by ESMF.
    The excited state SCF calculation took
    just two and a half times as long
    as the RHF calculation.
    \label{fig:pycm_orbs}
    }
\end{figure}

\subsection{PYCM}
\label{sec::pycm}

To verify that the benefits of the SCF approach are not
confined to smaller molecules, we exhibit its
use on a charge transfer state in the PYCM molecule 
that Subotnik used to demonstrate CIS's bias against
charge transfer states. \cite{Subotnik2011}
Working in a cc-pVDZ basis for the heavy atoms and
6-31G for hydrogen, we consider
the lowest charge transfer state,
for which we provide iteration-by-iteration
convergence details
in the Supplementary Material.
In Figure \ref{fig:pycm_orbs}, we plot
the ESMF prediction
for the donor and acceptor orbitals, which
in this case are just the relaxed HOMO and
LUMO orbitals as the $\bm{t}$ matrix
coefficients are strongly dominated by
the HOMO$\rightarrow$LUMO transition.
We see that this state transfers charge from
the $\pi$ bonding
orbital on the methylated ethylene moiety to
the $\pi^{*}$ orbital on the cyano-substituted
ethylene moiety.
Aside from the efficiency of the SCF solver in
this case (it takes just two-and-a-half times
as long as RHF when using the same Fock build code)
it is interesting to compare the prediction
against that of CIS, which is the analogous theory
when orbital relaxation is ignored.
CIS predicts a 7.30 eV excitation energy for
the lowest state in which this charge transfer
transition plays a significant role, whereas
ESMF predicts a 4.82 eV excitation energy.
This multiple-eV energy lowering after orbital
relaxation serves as a stark reminder of how
important these relaxations are for charge
transfer states.

%\textcolor{blue}{
%Aside from the core state (in which we anyways
%relied partly on the GVP), the states discussed
%above are all the lowest excitations, and so
%we would like to close with an example demonstrating
%that  the SCF approach to ESMF can also be used for
%higher excited states.
%}

\section{Conclusion}
\label{sec:conclusion}

In conclusion, orbital optimization in ESMF theory can
be formulated in terms of a one-electron
equation in which mean field operators provide
electron-electron repulsion and which is
brought to self-consistency
through an efficient iterative process that closely mirrors
ground state HF theory.
In particular, it is possible to formulate the excited
state many-electron energy in terms of four traces
between density matrices and mean field operators,
and the central commutator condition likewise contains
four commutators between these density matrices
and their partner mean field operators.
In a sense, this is a straightforward extension of
the HF case, where only one trace and one commutator
are needed.
As has long been true for Slater determinants, the SCF
approach to the ESMF orbitals appears to be significantly
more efficient than quasi-Newton methods, at least in
cases where the SCF iteration converges stably to the
desired state.
Looking forward, it will be interesting to see if,
as in the ground state case, the SCF approach admits
Kohn-Sham-style density functionals and whether
the optimization of the excitation coefficients can be
more tightly coupled to the optimization of the orbitals.

$\vspace{1mm}$

\noindent
{\small \textbf{SUPPLEMENTARY MATERIAL}}

$\vspace{1mm}$

See supplementary material for
additional mathematical details,
additional calculation details,
and molecular geometries.

$\vspace{1mm}$

%\begin{acknowledgments}

\noindent
{\small \textbf{ACKNOWLEDGEMENTS}}

$\vspace{1mm}$

%\noindent
%\textit{Acknowledgments} ---
This work was supported by the Early Career Research Program
of the Office of Science, Office of Basic Energy Sciences,
the U.S. Department of Energy, grant No.\ {DE-SC0017869}.
While final timing calculations were carried out on a
laptop, many preliminary calculations with earlier pilot
code were performed using the Berkeley Research Computing
Savio cluster.
%\end{acknowledgments}

$\vspace{1mm}$

\noindent
\textit{Data Availability Statement} ---
The data that supports the findings of this study
are available within the article
and its supplementary material.

%\bibliography{main}% Produces the bibliography via BibTeX.

\begin{thebibliography}{27}%
\makeatletter
\providecommand \@ifxundefined [1]{%
 \@ifx{#1\undefined}
}%
\providecommand \@ifnum [1]{%
 \ifnum #1\expandafter \@firstoftwo
 \else \expandafter \@secondoftwo
 \fi
}%
\providecommand \@ifx [1]{%
 \ifx #1\expandafter \@firstoftwo
 \else \expandafter \@secondoftwo
 \fi
}%
\providecommand \natexlab [1]{#1}%
\providecommand \enquote  [1]{``#1''}%
\providecommand \bibnamefont  [1]{#1}%
\providecommand \bibfnamefont [1]{#1}%
\providecommand \citenamefont [1]{#1}%
\providecommand \href@noop [0]{\@secondoftwo}%
\providecommand \href [0]{\begingroup \@sanitize@url \@href}%
\providecommand \@href[1]{\@@startlink{#1}\@@href}%
\providecommand \@@href[1]{\endgroup#1\@@endlink}%
\providecommand \@sanitize@url [0]{\catcode `\\12\catcode `\$12\catcode
  `\&12\catcode `\#12\catcode `\^12\catcode `\_12\catcode `\%12\relax}%
\providecommand \@@startlink[1]{}%
\providecommand \@@endlink[0]{}%
\providecommand \url  [0]{\begingroup\@sanitize@url \@url }%
\providecommand \@url [1]{\endgroup\@href {#1}{\urlprefix }}%
\providecommand \urlprefix  [0]{URL }%
\providecommand \Eprint [0]{\href }%
\providecommand \doibase [0]{http://dx.doi.org/}%
\providecommand \selectlanguage [0]{\@gobble}%
\providecommand \bibinfo  [0]{\@secondoftwo}%
\providecommand \bibfield  [0]{\@secondoftwo}%
\providecommand \translation [1]{[#1]}%
\providecommand \BibitemOpen [0]{}%
\providecommand \bibitemStop [0]{}%
\providecommand \bibitemNoStop [0]{.\EOS\space}%
\providecommand \EOS [0]{\spacefactor3000\relax}%
\providecommand \BibitemShut  [1]{\csname bibitem#1\endcsname}%
\let\auto@bib@innerbib\@empty
%</preamble>
\bibitem [{\citenamefont {Szabo}\ and\ \citenamefont
  {Ostlund}(1996)}]{Szabo-Ostland}%
  \BibitemOpen
  \bibfield  {author} {\bibinfo {author} {\bibfnamefont {A.}~\bibnamefont
  {Szabo}}\ and\ \bibinfo {author} {\bibfnamefont {N.~S.}\ \bibnamefont
  {Ostlund}},\ }\href@noop {} {\emph {\bibinfo {title} {Modern Quantum
  Chemistry: Introduction to Advanced Electronic Structure Theory}}}\ (\bibinfo
   {publisher} {Dover Publications},\ \bibinfo {address} {Mineola, N.Y.},\
  \bibinfo {year} {1996})\BibitemShut {NoStop}%
\bibitem [{\citenamefont {Helgaker}, \citenamefont {J{\o}gensen},\ and\
  \citenamefont {Olsen}(2000)}]{MolElecStruc}%
  \BibitemOpen
  \bibfield  {author} {\bibinfo {author} {\bibfnamefont {T.}~\bibnamefont
  {Helgaker}}, \bibinfo {author} {\bibfnamefont {P.}~\bibnamefont
  {J{\o}gensen}}, \ and\ \bibinfo {author} {\bibfnamefont {J.}~\bibnamefont
  {Olsen}},\ }\href@noop {} {\emph {\bibinfo {title} {Molecular Electronic
  Structure Theory}}}\ (\bibinfo  {publisher} {John Wiley and Sons, Ltd},\
  \bibinfo {address} {West Sussex, England},\ \bibinfo {year}
  {2000})\BibitemShut {NoStop}%
\bibitem [{\citenamefont {Shea}\ and\ \citenamefont
  {Neuscamman}(2018)}]{Shea2018}%
  \BibitemOpen
  \bibfield  {author} {\bibinfo {author} {\bibfnamefont {J.~A.~R.}\
  \bibnamefont {Shea}}\ and\ \bibinfo {author} {\bibfnamefont {E.}~\bibnamefont
  {Neuscamman}},\ }\bibfield  {title} {\enquote {\bibinfo {title}
  {{Communication: A mean field platform for excited state quantum
  chemistry}},}\ }\href@noop {} {\bibfield  {journal} {\bibinfo  {journal} {J.
  Chem. Phys.}\ }\textbf {\bibinfo {volume} {149}},\ \bibinfo {pages} {081101}
  (\bibinfo {year} {2018})}\BibitemShut {NoStop}%
\bibitem [{\citenamefont {Pulay}(1982)}]{pulay1982diis}%
  \BibitemOpen
  \bibfield  {author} {\bibinfo {author} {\bibfnamefont {P.}~\bibnamefont
  {Pulay}},\ }\bibfield  {title} {\enquote {\bibinfo {title} {Improved scf
  convergence acceleration},}\ }\href@noop {} {\bibfield  {journal} {\bibinfo
  {journal} {J. Comput. Chem.}\ }\textbf {\bibinfo {volume} {3}},\ \bibinfo
  {pages} {556--560} (\bibinfo {year} {1982})}\BibitemShut {NoStop}%
\bibitem [{\citenamefont {Zhao}\ and\ \citenamefont
  {Neuscamman}(2020{\natexlab{a}})}]{Zhao2019dft}%
  \BibitemOpen
  \bibfield  {author} {\bibinfo {author} {\bibfnamefont {L.}~\bibnamefont
  {Zhao}}\ and\ \bibinfo {author} {\bibfnamefont {E.}~\bibnamefont
  {Neuscamman}},\ }\bibfield  {title} {\enquote {\bibinfo {title} {Density
  functional extension to excited-state mean-field theory},}\ }\href@noop {}
  {\bibfield  {journal} {\bibinfo  {journal} {J. Chem. Theory Comput.}\
  }\textbf {\bibinfo {volume} {16}},\ \bibinfo {pages} {164} (\bibinfo {year}
  {2020}{\natexlab{a}})}\BibitemShut {NoStop}%
\bibitem [{\citenamefont {Shea}, \citenamefont {Gwin},\ and\ \citenamefont
  {Neuscamman}(2020)}]{Shea2020gvp}%
  \BibitemOpen
  \bibfield  {author} {\bibinfo {author} {\bibfnamefont {J.~A.~R.}\
  \bibnamefont {Shea}}, \bibinfo {author} {\bibfnamefont {E.}~\bibnamefont
  {Gwin}}, \ and\ \bibinfo {author} {\bibfnamefont {E.}~\bibnamefont
  {Neuscamman}},\ }\bibfield  {title} {\enquote {\bibinfo {title} {A
  generalized variational principle with applications to excited state mean
  field theory},}\ }\href@noop {} {\bibfield  {journal} {\bibinfo  {journal}
  {J. Chem. Theory Comput.}\ }\textbf {\bibinfo {volume} {16}},\ \bibinfo
  {pages} {1526} (\bibinfo {year} {2020})}\BibitemShut {NoStop}%
\bibitem [{\citenamefont {Clune}, \citenamefont {Shea},\ and\ \citenamefont
  {Neuscamman}(2020)}]{Clune2020topesmp2}%
  \BibitemOpen
  \bibfield  {author} {\bibinfo {author} {\bibfnamefont {R.}~\bibnamefont
  {Clune}}, \bibinfo {author} {\bibfnamefont {J.~A.~R.}\ \bibnamefont {Shea}},
  \ and\ \bibinfo {author} {\bibfnamefont {E.}~\bibnamefont {Neuscamman}},\
  }\bibfield  {title} {\enquote {\bibinfo {title} {An {N}$^5$-scaling
  excited-state-specific perturbation theory},}\ }\href@noop {} {\bibfield
  {journal} {\bibinfo  {journal} {arXiv}\ ,\ \bibinfo {pages} {2003.12923}}
  (\bibinfo {year} {2020})}\BibitemShut {NoStop}%
\bibitem [{\citenamefont {Zhao}\ and\ \citenamefont
  {Neuscamman}(2020{\natexlab{b}})}]{zhao2020esmf}%
  \BibitemOpen
  \bibfield  {author} {\bibinfo {author} {\bibfnamefont {L.}~\bibnamefont
  {Zhao}}\ and\ \bibinfo {author} {\bibfnamefont {E.}~\bibnamefont
  {Neuscamman}},\ }\bibfield  {title} {\enquote {\bibinfo {title} {Excited
  state mean-field theory without automatic differentiation},}\ }\href@noop {}
  {\bibfield  {journal} {\bibinfo  {journal} {J. Chem. Phys.}\ }\textbf
  {\bibinfo {volume} {152}},\ \bibinfo {pages} {204112} (\bibinfo {year}
  {2020}{\natexlab{b}})}\BibitemShut {NoStop}%
\bibitem [{\citenamefont {Bagus}(1965)}]{bagus1965scf}%
  \BibitemOpen
  \bibfield  {author} {\bibinfo {author} {\bibfnamefont {P.~S.}\ \bibnamefont
  {Bagus}},\ }\bibfield  {title} {\enquote {\bibinfo {title}
  {Self-consistent-field wave functions for hole states of some ne-like and
  ar-like ions},}\ }\href@noop {} {\bibfield  {journal} {\bibinfo  {journal}
  {Phys. Rev.}\ }\textbf {\bibinfo {volume} {139}},\ \bibinfo {pages} {A619}
  (\bibinfo {year} {1965})}\BibitemShut {NoStop}%
\bibitem [{\citenamefont {Hsu}, \citenamefont {Davidson},\ and\ \citenamefont
  {Pitzer}(1976)}]{Pitzer1976scf}%
  \BibitemOpen
  \bibfield  {author} {\bibinfo {author} {\bibfnamefont {H.-l.}\ \bibnamefont
  {Hsu}}, \bibinfo {author} {\bibfnamefont {E.~R.}\ \bibnamefont {Davidson}}, \
  and\ \bibinfo {author} {\bibfnamefont {R.~M.}\ \bibnamefont {Pitzer}},\
  }\bibfield  {title} {\enquote {\bibinfo {title} {An scf method for hole
  states},}\ }\href@noop {} {\bibfield  {journal} {\bibinfo  {journal} {J.
  Chem. Phys.}\ }\textbf {\bibinfo {volume} {65}},\ \bibinfo {pages} {609--613}
  (\bibinfo {year} {1976})}\BibitemShut {NoStop}%
\bibitem [{\citenamefont {Naves~de Brito}\ \emph {et~al.}(1991)\citenamefont
  {Naves~de Brito}, \citenamefont {Correia}, \citenamefont {Svensson},\ and\
  \citenamefont {{\AA}gren}}]{argen1991xray}%
  \BibitemOpen
  \bibfield  {author} {\bibinfo {author} {\bibfnamefont {A.}~\bibnamefont
  {Naves~de Brito}}, \bibinfo {author} {\bibfnamefont {N.}~\bibnamefont
  {Correia}}, \bibinfo {author} {\bibfnamefont {S.}~\bibnamefont {Svensson}}, \
  and\ \bibinfo {author} {\bibfnamefont {H.}~\bibnamefont {{\AA}gren}},\
  }\bibfield  {title} {\enquote {\bibinfo {title} {A theoretical study of x-ray
  photoelectron spectra of model molecules for polymethylmethacrylate},}\
  }\href@noop {} {\bibfield  {journal} {\bibinfo  {journal} {J. Chem. Phys.}\
  }\textbf {\bibinfo {volume} {95}},\ \bibinfo {pages} {2965--2974} (\bibinfo
  {year} {1991})}\BibitemShut {NoStop}%
\bibitem [{\citenamefont {Besley}, \citenamefont {Gilbert},\ and\ \citenamefont
  {Gill}(2009)}]{Gill2009dscf}%
  \BibitemOpen
  \bibfield  {author} {\bibinfo {author} {\bibfnamefont {N.~A.}\ \bibnamefont
  {Besley}}, \bibinfo {author} {\bibfnamefont {A.~T.}\ \bibnamefont {Gilbert}},
  \ and\ \bibinfo {author} {\bibfnamefont {P.~M.}\ \bibnamefont {Gill}},\
  }\bibfield  {title} {\enquote {\bibinfo {title} {Self-consistent-field
  calculations of core excited states},}\ }\href@noop {} {\bibfield  {journal}
  {\bibinfo  {journal} {J. Chem. Phys.}\ }\textbf {\bibinfo {volume} {130}},\
  \bibinfo {pages} {124308} (\bibinfo {year} {2009})}\BibitemShut {NoStop}%
\bibitem [{\citenamefont {Filatov}\ and\ \citenamefont
  {Shaik}(1999)}]{Shaik1999}%
  \BibitemOpen
  \bibfield  {author} {\bibinfo {author} {\bibfnamefont {M.}~\bibnamefont
  {Filatov}}\ and\ \bibinfo {author} {\bibfnamefont {S.}~\bibnamefont
  {Shaik}},\ }\bibfield  {title} {\enquote {\bibinfo {title} {A spin-restricted
  ensemble-referenced kohn-sham method and its application to diradicaloid
  situations},}\ }\href@noop {} {\bibfield  {journal} {\bibinfo  {journal}
  {Chem. Phys. Lett.}\ }\textbf {\bibinfo {volume} {304}},\ \bibinfo {pages}
  {429--437} (\bibinfo {year} {1999})}\BibitemShut {NoStop}%
\bibitem [{\citenamefont {Kowalczyk}\ \emph {et~al.}(2013)\citenamefont
  {Kowalczyk}, \citenamefont {Tsuchimochi}, \citenamefont {Chen}, \citenamefont
  {Top},\ and\ \citenamefont {{Van Voorhis}}}]{Kowalczyk2013}%
  \BibitemOpen
  \bibfield  {author} {\bibinfo {author} {\bibfnamefont {T.}~\bibnamefont
  {Kowalczyk}}, \bibinfo {author} {\bibfnamefont {T.}~\bibnamefont
  {Tsuchimochi}}, \bibinfo {author} {\bibfnamefont {P.~T.}\ \bibnamefont
  {Chen}}, \bibinfo {author} {\bibfnamefont {L.}~\bibnamefont {Top}}, \ and\
  \bibinfo {author} {\bibfnamefont {T.}~\bibnamefont {{Van Voorhis}}},\
  }\bibfield  {title} {\enquote {\bibinfo {title} {{Excitation energies and
  Stokes shifts from a restricted open-shell Kohn-Sham approach}},}\
  }\href@noop {} {\bibfield  {journal} {\bibinfo  {journal} {J. Chem. Phys.}\
  }\textbf {\bibinfo {volume} {138}} (\bibinfo {year} {2013})}\BibitemShut
  {NoStop}%
\bibitem [{\citenamefont {Wu}\ and\ \citenamefont
  {Van~Voorhis}(2005)}]{VanVoorhis2005cdft}%
  \BibitemOpen
  \bibfield  {author} {\bibinfo {author} {\bibfnamefont {Q.}~\bibnamefont
  {Wu}}\ and\ \bibinfo {author} {\bibfnamefont {T.}~\bibnamefont
  {Van~Voorhis}},\ }\bibfield  {title} {\enquote {\bibinfo {title} {Direct
  optimization method to study constrained systems within density-functional
  theory},}\ }\href@noop {} {\bibfield  {journal} {\bibinfo  {journal} {Phys.
  Rev. A}\ }\textbf {\bibinfo {volume} {72}},\ \bibinfo {pages} {024502}
  (\bibinfo {year} {2005})}\BibitemShut {NoStop}%
\bibitem [{\citenamefont {Theophilou}(1979)}]{theophilou1979}%
  \BibitemOpen
  \bibfield  {author} {\bibinfo {author} {\bibfnamefont {A.~K.}\ \bibnamefont
  {Theophilou}},\ }\bibfield  {title} {\enquote {\bibinfo {title} {The energy
  density functional formalism for excited states},}\ }\href@noop {} {\bibfield
   {journal} {\bibinfo  {journal} {J. Phys. C}\ }\textbf {\bibinfo {volume}
  {12}},\ \bibinfo {pages} {5419} (\bibinfo {year} {1979})}\BibitemShut
  {NoStop}%
\bibitem [{\citenamefont {Kohn}(1986)}]{kohn1986quailocal}%
  \BibitemOpen
  \bibfield  {author} {\bibinfo {author} {\bibfnamefont {W.}~\bibnamefont
  {Kohn}},\ }\bibfield  {title} {\enquote {\bibinfo {title} {Density-functional
  theory for excited states in a quasi-local-density approximation},}\
  }\href@noop {} {\bibfield  {journal} {\bibinfo  {journal} {Phys. Rev. A}\
  }\textbf {\bibinfo {volume} {34}},\ \bibinfo {pages} {737} (\bibinfo {year}
  {1986})}\BibitemShut {NoStop}%
\bibitem [{\citenamefont {Gross}, \citenamefont {Oliveira},\ and\ \citenamefont
  {Kohn}(1988)}]{gross1988density}%
  \BibitemOpen
  \bibfield  {author} {\bibinfo {author} {\bibfnamefont {E.~K.}\ \bibnamefont
  {Gross}}, \bibinfo {author} {\bibfnamefont {L.~N.}\ \bibnamefont {Oliveira}},
  \ and\ \bibinfo {author} {\bibfnamefont {W.}~\bibnamefont {Kohn}},\
  }\bibfield  {title} {\enquote {\bibinfo {title} {Density-functional theory
  for ensembles of fractionally occupied states. i. basic formalism},}\
  }\href@noop {} {\bibfield  {journal} {\bibinfo  {journal} {Phys. Rev. A}\
  }\textbf {\bibinfo {volume} {37}},\ \bibinfo {pages} {2809} (\bibinfo {year}
  {1988})}\BibitemShut {NoStop}%
\bibitem [{\citenamefont {Jim{\'e}nez-Hoyos}\ \emph {et~al.}(2012)\citenamefont
  {Jim{\'e}nez-Hoyos}, \citenamefont {Henderson}, \citenamefont {Tsuchimochi},\
  and\ \citenamefont {Scuseria}}]{jimenez2012PHF}%
  \BibitemOpen
  \bibfield  {author} {\bibinfo {author} {\bibfnamefont {C.~A.}\ \bibnamefont
  {Jim{\'e}nez-Hoyos}}, \bibinfo {author} {\bibfnamefont {T.~M.}\ \bibnamefont
  {Henderson}}, \bibinfo {author} {\bibfnamefont {T.}~\bibnamefont
  {Tsuchimochi}}, \ and\ \bibinfo {author} {\bibfnamefont {G.~E.}\ \bibnamefont
  {Scuseria}},\ }\bibfield  {title} {\enquote {\bibinfo {title} {Projected
  hartree--fock theory},}\ }\href@noop {} {\bibfield  {journal} {\bibinfo
  {journal} {J. Chem. Phys.}\ }\textbf {\bibinfo {volume} {136}},\ \bibinfo
  {pages} {164109} (\bibinfo {year} {2012})}\BibitemShut {NoStop}%
\bibitem [{\citenamefont {Ye}\ \emph {et~al.}(2017)\citenamefont {Ye},
  \citenamefont {Welborn}, \citenamefont {Ricke},\ and\ \citenamefont
  {Van~Voorhis}}]{vanVoorhis2017sigmaSCF}%
  \BibitemOpen
  \bibfield  {author} {\bibinfo {author} {\bibfnamefont {H.-Z.}\ \bibnamefont
  {Ye}}, \bibinfo {author} {\bibfnamefont {M.}~\bibnamefont {Welborn}},
  \bibinfo {author} {\bibfnamefont {N.~D.}\ \bibnamefont {Ricke}}, \ and\
  \bibinfo {author} {\bibfnamefont {T.}~\bibnamefont {Van~Voorhis}},\
  }\bibfield  {title} {\enquote {\bibinfo {title} {$\sigma$-scf: A direct
  energy-targeting method to mean-field excited states},}\ }\href@noop {}
  {\bibfield  {journal} {\bibinfo  {journal} {J. Chem. Phys}\ }\textbf
  {\bibinfo {volume} {147}},\ \bibinfo {pages} {214104} (\bibinfo {year}
  {2017})}\BibitemShut {NoStop}%
\bibitem [{\citenamefont {Ye}\ and\ \citenamefont
  {Voorhis}(2019)}]{Voorhis2019}%
  \BibitemOpen
  \bibfield  {author} {\bibinfo {author} {\bibfnamefont {H.}~\bibnamefont
  {Ye}}\ and\ \bibinfo {author} {\bibfnamefont {T.~V.}\ \bibnamefont
  {Voorhis}},\ }\bibfield  {title} {\enquote {\bibinfo {title} {Half-projected
  $\sigma$ self-consistent field for electronic excited states},}\ }\href@noop
  {} {\bibfield  {journal} {\bibinfo  {journal} {J. Chem. Theory Comput.}\
  }\textbf {\bibinfo {volume} {15(5)}},\ \bibinfo {pages} {2954--2964}
  (\bibinfo {year} {2019})}\BibitemShut {NoStop}%
\bibitem [{\citenamefont {Hait}\ and\ \citenamefont
  {Head-Gordon}(2020)}]{hait2020oo}%
  \BibitemOpen
  \bibfield  {author} {\bibinfo {author} {\bibfnamefont {D.}~\bibnamefont
  {Hait}}\ and\ \bibinfo {author} {\bibfnamefont {M.}~\bibnamefont
  {Head-Gordon}},\ }\bibfield  {title} {\enquote {\bibinfo {title} {Excited
  state orbital optimization via minimizing the square of the gradient: General
  approach and application to singly and doubly excited states via density
  functional theory},}\ }\href@noop {} {\bibfield  {journal} {\bibinfo
  {journal} {J. Chem. Theory Comput.}\ }\textbf {\bibinfo {volume} {16}},\
  \bibinfo {pages} {1699--1710} (\bibinfo {year} {2020})}\BibitemShut {NoStop}%
\bibitem [{\citenamefont {Van~Voorhis}\ and\ \citenamefont
  {Head-Gordon}(2002)}]{VanVoorhis2002gdm}%
  \BibitemOpen
  \bibfield  {author} {\bibinfo {author} {\bibfnamefont {T.}~\bibnamefont
  {Van~Voorhis}}\ and\ \bibinfo {author} {\bibfnamefont {M.}~\bibnamefont
  {Head-Gordon}},\ }\bibfield  {title} {\enquote {\bibinfo {title} {A geometric
  approach to direct minimization},}\ }\href@noop {} {\bibfield  {journal}
  {\bibinfo  {journal} {Mol. Phys.}\ }\textbf {\bibinfo {volume} {100}},\
  \bibinfo {pages} {1713--1721} (\bibinfo {year} {2002})}\BibitemShut {NoStop}%
\bibitem [{\citenamefont {McWeeny}(1996)}]{McWeeny}%
  \BibitemOpen
  \bibfield  {author} {\bibinfo {author} {\bibfnamefont {R.}~\bibnamefont
  {McWeeny}},\ }\href@noop {} {\emph {\bibinfo {title} {Methods of Molecular
  Quantum Mechanics}}}\ (\bibinfo  {publisher} {Academic Press},\ \bibinfo
  {address} {London},\ \bibinfo {year} {1996})\BibitemShut {NoStop}%
\bibitem [{\citenamefont {Kreplin}, \citenamefont {Knowles},\ and\
  \citenamefont {Werner}(2020)}]{kreplin2020mcscf}%
  \BibitemOpen
  \bibfield  {author} {\bibinfo {author} {\bibfnamefont {D.~A.}\ \bibnamefont
  {Kreplin}}, \bibinfo {author} {\bibfnamefont {P.~J.}\ \bibnamefont
  {Knowles}}, \ and\ \bibinfo {author} {\bibfnamefont {H.-J.}\ \bibnamefont
  {Werner}},\ }\bibfield  {title} {\enquote {\bibinfo {title} {Mcscf
  optimization revisited. ii. combined first-and second-order orbital
  optimization for large molecules},}\ }\href@noop {} {\bibfield  {journal}
  {\bibinfo  {journal} {J. Chem. Phys.}\ }\textbf {\bibinfo {volume} {152}},\
  \bibinfo {pages} {074102} (\bibinfo {year} {2020})}\BibitemShut {NoStop}%
\bibitem [{\citenamefont {Schirmer}\ \emph {et~al.}(1993)\citenamefont
  {Schirmer}, \citenamefont {Trofimov}, \citenamefont {Randall}, \citenamefont
  {Feldhaus}, \citenamefont {Bradshaw}, \citenamefont {Ma}, \citenamefont
  {Chen},\ and\ \citenamefont {Sette}}]{schirmer1993}%
  \BibitemOpen
  \bibfield  {author} {\bibinfo {author} {\bibfnamefont {J.}~\bibnamefont
  {Schirmer}}, \bibinfo {author} {\bibfnamefont {A.~B.}\ \bibnamefont
  {Trofimov}}, \bibinfo {author} {\bibfnamefont {K.~J.}\ \bibnamefont
  {Randall}}, \bibinfo {author} {\bibfnamefont {J.}~\bibnamefont {Feldhaus}},
  \bibinfo {author} {\bibfnamefont {A.~M.}\ \bibnamefont {Bradshaw}}, \bibinfo
  {author} {\bibfnamefont {Y.}~\bibnamefont {Ma}}, \bibinfo {author}
  {\bibfnamefont {C.~T.}\ \bibnamefont {Chen}}, \ and\ \bibinfo {author}
  {\bibfnamefont {F.}~\bibnamefont {Sette}},\ }\bibfield  {title} {\enquote
  {\bibinfo {title} {K-shell excitation of the water, ammonia, and methane
  molecules using high-resolution photoabsorption spectroscopy},}\ }\href@noop
  {} {\bibfield  {journal} {\bibinfo  {journal} {Phys. Rev. A}\ }\textbf
  {\bibinfo {volume} {47}},\ \bibinfo {pages} {1136} (\bibinfo {year}
  {1993})}\BibitemShut {NoStop}%
\bibitem [{\citenamefont {Subotnik}(2011)}]{Subotnik2011}%
  \BibitemOpen
  \bibfield  {author} {\bibinfo {author} {\bibfnamefont {J.~E.}\ \bibnamefont
  {Subotnik}},\ }\bibfield  {title} {\enquote {\bibinfo {title} {Communication:
  Configuration interaction singles has a large systematic bias against
  charge-transfer states},}\ }\href@noop {} {\bibfield  {journal} {\bibinfo
  {journal} {J. Chem. Phys.}\ }\textbf {\bibinfo {volume} {135}},\ \bibinfo
  {pages} {071104} (\bibinfo {year} {2011})}\BibitemShut {NoStop}%
\end{thebibliography}
%

\clearpage

\appendix
\renewcommand\theequation{S\arabic{equation}}
\setcounter{equation}{0}

\renewcommand{\thetable}{S\arabic{table}}
\setcounter{table}{0}
\section{Mathematical Detail}

First, let's rewrite the ESMF ansatz in Eq.\ (\ref{eqn:esmf_ansatz}) as
\begin{equation}
| \Psi_{ESMF} \rangle =
\sum_{ia} t_{ia} a^\dagger_{\uparrow  } i_{\uparrow  } | \Phi_A \rangle
        + t_{ia} a^\dagger_{\downarrow} i_{\downarrow} | \Phi_A \rangle
\label{eqn:esmf_wfn}
\end{equation}
where $\Phi_A$ refers to the Aufbau determinant and
$a^\dagger_\uparrow$ and $i^\dagger_\uparrow$ are
$\uparrow$-spin creation operators for virtual and
occupied orbitals, respectively, in the ESMF molecular
orbital basis.
We define the Hamiltonian in chemist's notation as
%\textcolor{red}{Shouldn't there be a 1/2 in front of the (pq|rs) term?}
\begin{align}
\hat{H} = & \hphantom{+}
\sum_{pq} h^{(MO)}_{pq}
\big( \hspace{0.5mm}
p^\dagger_\uparrow   q_\uparrow
\hspace{0.5mm} + \hspace{0.5mm}
p^\dagger_\downarrow q_\downarrow
\hspace{0.5mm} \big)
\notag \\
& + \frac{1}{2}\sum_{pqrs} \{pq|rs\}
\big( \hspace{1mm}
p^\dagger_\uparrow   r^\dagger_\uparrow   s_\uparrow   q_\uparrow   
\hspace{0.5mm} + \hspace{0.5mm}
p^\dagger_\downarrow r^\dagger_\downarrow s_\downarrow q_\downarrow
\notag \\
& \hspace{24mm} + \hspace{0.5mm}
p^\dagger_\uparrow   r^\dagger_\downarrow s_\downarrow q_\uparrow 
\hspace{0.5mm} + \hspace{0.5mm}
p^\dagger_\downarrow r^\dagger_\uparrow   s_\uparrow   q_\downarrow   
\hspace{1mm} \big)
\label{eqn:ham}
\end{align}
where $\bm{h}^{(MO)}$ and $\{pq|rs\}$
are the one- and two-electron
integrals in the ESMF orbital basis.
The ESMF energy is
\begin{align}
 E_{ESMF} = \langle  \Psi_{ESMF} |\hat{H}  |\Psi_{ESMF}\rangle.
 \label{eqn:esmf_e}
\end{align}
From here on out, we adopt a 
summation convention in which a sum is implied
over any index that appears more than once
in the same term.
Using this convention, and working through the
second-quantized algebra,
the ESMF energy becomes
\begin{align}
E_{ESMF} = & \hphantom{+}
   2 \hspace{1mm} t_{ia} t_{ib} h^{(MO)}_{ab}
 - 2 \hspace{1mm} t_{ia} t_{ja} h^{(MO)}_{ij}
 + 4 \hspace{1mm} t_{ia} t_{ia} h^{(MO)}_{kk}
\notag \\[1mm]
%%%%
&+ 4 \hspace{1mm} t_{ia} t_{ib} \{ab|kk\}
 - 2 \hspace{1mm} t_{ia} t_{ib} \{ak|bk\}
\notag \\[1mm]
%%%%
&- 4 \hspace{1mm} t_{ia} t_{ja} \{ij|kk\}
 + 2 \hspace{1mm} t_{ia} t_{ja} \{ik|jk\}
\notag \\[1mm]
%%%%
&+ 4 \hspace{1mm} t_{ia} t_{ia} \{kk|ll\}
 - 2 \hspace{1mm} t_{ia} t_{ia} \{kl|kl\}
\notag \\[1mm]
%%%%
&+ 4 \hspace{1mm} t_{ia} t_{jb} \{ia|jb\}
 - 2 \hspace{1mm} t_{ia} t_{jb} \{ij|ab\}
\label{eqn:sqa_energy}
%E_{ESMF} = \hphantom{+} &
%\uline{2 \sum_{ab} t_{ia} t_{ib} h_{ab}}
%- \uline{2 \sum_{ij} t_{ia} t_{ja} h_{ij}} \notag \\
%&+ \uline{4 \sum_{k}t_{ia} t_{ia} h_{kk}}
%+ \uline{8 \sum_{abk} t_{ia} t_{ib} (ab | kk)} \notag \\
%& - \uline{8 \sum_{ijk}t_{ia} t_{ja} (ij|kk)}
%+  \uline{8 \sum_{kl}t_{ia} t_{ia}(kk|ll)} \notag \\
%& + 8\sum_{aijb} t_{ia}t_{jb} (ai|jb)
%- \uline{4 \sum_{abk} t_{ia} t_{ib} (kb|ak)} \notag \\
%&+ \uline{4 \sum_{ijk} t_{ia} t_{ja} (ki | jk)}
%- \uline{4 \sum_{kl} t_{ia} t_{ia} (kl|kl)}
%- 4 \sum_{abij} t_{ia} t_{jb} (ab|ij)
\end{align}
%where summation is implied for the
%repeated indices in each term and
where $a, b, c, d$ are virtual orbitals,
$i, j, k, l, m$ are occupied orbitals,
and we will use $p, q, r, s, w, x, y, z$
for general orbitals.
Note the pattern in the terms: first, the one electron integrals
are contracted and summed over,
then the two electron integrals.
The two electron integral terms come in pairs,
one part of each pair corresponding to the conventional coulomb operator
and the other to the conventional exchange operator.
As in RHF theory, the coulomb terms have an extra factor of two
compared to the exchange terms, which is what produces the
general pattern of $2J - K$ in the final mean field operators. 

To derive Eq.\ (\ref{eqn:esmf_roothaan}), we start with the
Lagrangian in Eq.\ (\ref{eqn:esmf_lagrangian}), take
derivatives with respect to the elements of the
orbital coefficient matrix $\bm{C}$,
%unitary
%orbital rotation matrix $\bm{U}$,
%\begin{align}
%    \bm{U} = \exp (\bm{X}),
%\end{align}
and then set these derivatives equal to zero.
Derivatives of the Lagrange multiplier term in
Eq.\ (\ref{eqn:esmf_lagrangian}) lead to the right
hand side of Eq.\ (\ref{eqn:esmf_roothaan}) in the same
way that they do for the HF Roothaan equation, and so we will
not work through them explicitly.
For the derivatives of the ESMF energy, we need to remember
that the one- and two-electron integrals in the ESMF
molecular orbital basis are transformed from the 
atomic orbital basis as
\begin{align}
    h^{(MO)}_{pq}  &= C_{rp} h_{rs} C_{sq} \\
    \{pq|rs\} &= C_{wp} C_{xq} C_{yr} C_{zs} (wx|yz)
\end{align}
where $\bm{h}$ and $(wx|yz)$ are the one- and
two-electron integrals (again in chemists' notation)
in the atomic orbital basis.
Noting that $\bm{C}$ contains an ``occupied'' block
$C^{(o)}$ (the first $n_{o}$ columns)
and a ``virtual'' block $C^{(v)}$
(the remaining columns),
we can consider the derivatives with respect to the
elements of these blocks separately.
As an example, the occupied block derivative of the
first two-electron integral term in
Eq.\ (\ref{eqn:sqa_energy}) is
(here $m$ is an occupied index and $x$ is
a general index)
\begin{align}
\frac{\partial}{\partial C_{xm}}
  \bigg( t_{ia} t_{ib} \{ab|kk\} \bigg)
% = & \hphantom{+}
%   t_{ia} t_{ib} C_{pa} C_{qb} C_{sm} \{pq |xs\} \notag \\
%&+ t_{ia} t_{ib} C_{pa} C_{qb} C_{rm} \{pq |rx\} \\[2mm]
 = & \hphantom{+}
   2 \hspace{1mm}
   t_{ia} t_{ib} C_{pa} C_{qb} C_{sm} (pq|xs).
   \label{eqn:example_occ}
\end{align}
The analogous virtual block derivative is
\begin{align}
\frac{\partial}{\partial C_{xc}}
  \bigg( t_{ia} t_{ib} \{ab|kk\} \bigg)
 = 2 \hspace{1mm}
 t_{ic} t_{ia} C_{qa} C_{rk} C_{sk} (xq|rs).
   \label{eqn:example_vir}
\end{align}
%Again, these formulae rely on the repeated-index
%summation convention.
We now define
\begin{align}
\mathcal{A}^{coul}_{pq} & = \hspace{1mm}
C_{rk} C_{sk} (pq|rs) \\[2mm]
\mathcal{B}^{coul}_{pq} & = \hspace{1mm}
C_{sb} t_{ib} t_{ia} C_{ra} (pq|rs) \\[2mm]
\mathcal{C}^{coul}_{pq} & = \hspace{1mm}
 C_{ri} t_{ia} t_{ja} C_{sj} (pq|rs) \\[2mm]
 \mathcal{D}^{coul}_{pq} & = \hspace{1mm}
 C_{ri} t_{ia} C_{sa} (pq|rs)
\end{align}
%\textcolor{red}{why did we drop the dagger/transposes when defining A/B/C? easier to follow with them, right?}
% Oh, wait, I see what you mean.
% Personally I find it easier to follow without the transposes...
as well as
$\mathcal{A}^{exch}$,
$\mathcal{B}^{exch}$, 
$\mathcal{C}^{exch}$, and
$\mathcal{D}^{exch}$,
which are the same except for having
$q$ and $r$ swapped in the two-electron integral.
With these definitions, we can combine 
Eqs.\ (\ref{eqn:example_occ}) and (\ref{eqn:example_vir})
and write the derivative in matrix form.
\begin{align}
\frac{\partial}{\partial \bm{C} }
  \bigg( t_{ia} t_{ib} \{ab|kk\} \bigg)
 =
&
 \bigg[ \hspace{1mm}
    2 \hspace{1mm} \mathcal{B}^{coul} \bm{C}^{(o)}
 \hspace{2mm} \bigg| \hspace{2mm}
    \bm{0}
 \hspace{1mm} \bigg] \notag \\[1mm]
& \hspace{2mm} + 
 \bigg[ \hspace{1mm}
    \bm{0}
 \hspace{2mm} \bigg| \hspace{2mm}
    2 \hspace{1mm} \mathcal{A}^{coul} \bm{C}^{(v)} \bm{t}^T \bm{t}
 \hspace{1mm} \bigg]
 %\notag \\
 %\quad \hspace{2} \hphantom{a}
 \label{eqn:example_mat_deriv}
\end{align}
where on the right hand side we have placed a vertical
bar to separate the occupied and virtual blocks of
the matrices.
Note that we have kept the two matrices on the right
hand side here separate, as they will end up contributing
to different terms within
Eq.\ (\ref{eqn:esmf_roothaan}), which we enumerate as
follows.
%Speaking of the different terms in the left hand side of
%Eq.\ (\ref{eqn:esmf_roothaan}) that we wish to derive,
%allow us to enumerate them.
\begin{align}
\mathrm{Term~1} & \quad \big( \bm{h}+\bm{W}[\bm{A}] \big) C \gamma^{(MO)}
\\
\mathrm{Term~2} & \quad \bm{W}[\bm{D}] C A^{(MO)}
\\
\mathrm{Term~3} & \quad \bm{W}[\bm{T}] C (T^{(MO)})^T
\\
\mathrm{Term~4} & \quad (\bm{W}[\bm{T}])^T C T^{(MO)}
\end{align}
We will explicitly work through to Terms 1 and 2.
Terms 3 and 4 are derived similarly.
Now, each of the first nine terms in 
Eq.\ (\ref{eqn:sqa_energy}) makes a contribution
to Terms 1 or 2.
We have already worked out the $\bm{C}$-derivatives
for one of these terms (the fourth one) in
Eq.\ (\ref{eqn:example_mat_deriv}).
All the others are listed here. Note that the last two derivatives do not make a contribution to Terms 1 or 2, but are listed for completeness. 
%Note we do not list the derivatives for the
%last two terms from Eq.\ (\ref{eqn:sqa_energy})
%as they do not make a contribution to Term 1.
%\textcolor{red}{Should we just list the last two terms anyway for completeness?}
%\textcolor{red}{YES!}
\begin{align*}
\frac{\partial}{\partial \bm{C} }
  \bigg( t_{ia} t_{ib} h^{(MO)}_{ab} \bigg)
 =
& \hspace{1mm}
 \bigg[ \hspace{1mm}
    \bm{0}
 \hspace{2mm} \bigg| \hspace{2mm}
    2 \hspace{1mm} \bm{h} \hspace{0.5mm}
    \bm{C}^{(v)} \bm{t}^T \bm{t}
 \hspace{1mm} \bigg]
 \notag \\[2mm]
 %\quad \hspace{2} \hphantom{a}
%\end{align*}
%
%\begin{align*}
    \frac{\partial}{\partial \bm{C} }
  \bigg( t_{ia} t_{ja} h^{(MO)}_{ij} \bigg)
 = 
 & \hspace{1mm}
 \bigg [ \hspace{1mm}
    2 \hspace{1mm} \bm{h} \hspace{0.5mm}
    \bm{C}^{(o)} \bm{t} \hspace{0.5mm} \bm{t}^T 
 \hspace{2mm} \bigg| \hspace{2mm}
    \bm{0}
 \hspace{1mm} \bigg]
 \notag \\[2mm]
%\quad \hspace{2} \hphantom{a}
%\end{align*}
%
%\begin{align*}
    \frac{\partial}{\partial \bm{C} }
  \bigg( t_{ia} t_{ia} h^{(MO)}_{kk} \bigg)
 = 
 & \hspace{1mm}
 \bigg [ \hspace{1mm}
     \hspace{1mm} \bm{h} \hspace{0.5mm}
     \hspace{1mm} \bm{C}^{(o)} 
 \hspace{2mm} \bigg| \hspace{2mm}
    \bm{0}
 \hspace{1mm} \bigg]
 \notag \\[2mm]
%\quad \hspace{2} \hphantom{a}
%\end{align*}
%
%\begin{align*}
\frac{\partial}{\partial \bm{C} }
  \bigg( t_{ia} t_{ib} \{ak|bk\} \bigg)
 =
& \hspace{1mm}
 \bigg[ \hspace{1mm}
    2 \hspace{1mm} \mathcal{B}^{exch} \bm{C}^{(o)}
 \hspace{2mm} \bigg| \hspace{2mm}
    \bm{0}
 \hspace{1mm} \bigg] \notag \\[1mm]
& \hspace{4mm} + \hspace{1mm}
 \bigg[ \hspace{1mm}
    \bm{0}
 \hspace{2mm} \bigg| \hspace{2mm}
    2 \hspace{1mm} \mathcal{A}^{exch} \bm{C}^{(v)} \bm{t}^T \bm{t}
 \hspace{1mm} \bigg]
 \notag \\[2mm]
 %\quad \hspace{2} \hphantom{a}
%\end{align*}
%
%\begin{align*}
    \frac{\partial}{\partial \bm{C} }
  \bigg( t_{ia} t_{ja} \{ij|kk\} \bigg)
 = 
 & \hspace{1mm}
 \bigg [ \hspace{1mm}
    2 \hspace{1mm} \mathcal{C}^{coul} \bm{C}^{(o)}
 \hspace{2mm} \bigg| \hspace{2mm}
    \bm{0}
 \hspace{1mm} \bigg] \notag \\[1mm]
& \hspace{4mm} + 
 \bigg[ \hspace{1mm}
    2 \hspace{1mm} \mathcal{A}^{coul} \bm{C}^{(o)}
    \bm{t} \hspace{0.5mm} \bm{t}^T 
 \hspace{2mm} 
 \bigg| \hspace{2mm} \bm{0} \hspace{1mm} \bigg]
 \notag \\[2mm]
%\quad \hspace{2} \hphantom{a}
%\end{align*}
% 
%\begin{align*}
    \frac{\partial}{\partial \bm{C} }
  \bigg( t_{ia} t_{ja} \{ik|jk\} \bigg)
 = 
 & \hspace{1mm}
 \bigg [ \hspace{1mm}
    2 \hspace{1mm} \mathcal{C}^{exch} \bm{C}^{(o)}
 \hspace{2mm} \bigg| \hspace{2mm}
    \bm{0}
 \hspace{1mm} \bigg] \notag \\[1mm]
& \hspace{4mm} + 
 \bigg[ \hspace{1mm}
    2 \hspace{1mm} \mathcal{A}^{exch} \bm{C}^{(o)}
    \bm{t} \hspace{0.5mm} \bm{t}^T 
 \hspace{2mm} 
 \bigg| \hspace{2mm} \bm{0} \hspace{1mm} \bigg]
 \notag \\[2mm]
%\quad \hspace{2} \hphantom{a}
%\end{align*}
%
%\begin{align*}
    \frac{\partial}{\partial \bm{C} }
  \bigg( t_{ia} t_{ia} \{kk|ll\} \bigg)
 = 
 & \hspace{1mm}
 \bigg [ \hspace{1mm}
    2 \hspace{1mm} \mathcal{A}^{coul} \bm{C}^{(o)}
 \hspace{2mm} \bigg| \hspace{2mm}
    \bm{0}
 \hspace{1mm} \bigg]
 \notag \\[2mm]
%\quad \hspace{2} \hphantom{a}
%\end{align*}
%
%\begin{align*}
    \frac{\partial}{\partial \bm{C} }
  \bigg( t_{ia} t_{ia} \{kl|kl\} \bigg)
 = 
 & \hspace{1mm}
 \bigg [ \hspace{1mm}
    2 \hspace{1mm} \mathcal{A}^{exch} \bm{C}^{(o)}
 \hspace{2mm} \bigg| \hspace{2mm}
    \bm{0}
 \hspace{1mm} \bigg]
 \notag \\[2mm]
 %\notag \\[1mm]
%\quad \hspace{2} \hphantom{a}
    \frac{\partial}{\partial \bm{C} }
  \bigg( t_{ia} t_{jb} \{ia|jb\} \bigg)
 = 
 & \hspace{1mm}
 \bigg [ \hspace{1mm}
    2 \hspace{1mm} \mathcal{D}^{coul} \bm{C}^{(o)} \bm{t}
 \hspace{3.75mm} \bigg| \hspace{2mm}
   2 \hspace{1mm} \mathcal{D}^{coul} \bm{C}^{(v)}
    \hspace{0.5mm} \bm{t}^T 
 \hspace{1mm} \bigg] 
 \notag \\[2mm]
 %\notag \\[1mm]
%\quad \hspace{2} \hphantom{a}
    \frac{\partial}{\partial \bm{C} }
  \bigg( t_{ia} t_{jb} \{ij|ab\} \bigg)
 = 
 & \hspace{1mm}
 \bigg [ \hspace{1mm}
    2 \hspace{1mm} \mathcal{D}^{exch^T} \bm{C}^{(o)} \bm{t}
 \hspace{2mm} \bigg| \hspace{2mm}
   2 \hspace{1mm} \mathcal{D}^{exch} \bm{C}^{(v)}
    \hspace{0.5mm} \bm{t}^T 
 \hspace{1mm} \bigg] 
\end{align*}
%%%
Using these $\bm{C}$-derivatives, we get to
Term 1 by adding up all the pieces from
the derivatives of the terms
in Eq.\ (\ref{eqn:sqa_energy}) that
involve $\bm{h}$, 
$\mathcal{A}^{coul}$,
or  $\mathcal{A}^{exch}$.
If we make the definition
$\mathcal{A}=2\mathcal{A}^{coul}-\mathcal{A}^{exch}$,
then this addition comes out to
\begin{align}
 4 \hspace{1mm} \bigg [ \hspace{1mm}
    ( \bm{h} + \mathcal{A} ) \bm{C}^{(o)}
    ( \bm{I}^{(o)} - \bm{t} \hspace{0.5mm} \bm{t}^T )
 \hspace{2mm} \bigg| \hspace{2mm}
    ( \bm{h} + \mathcal{A} ) \bm{C}^{(v)}
    \bm{t}^T \bm{t}
 \hspace{1mm} \bigg]
\end{align}
where $I^{(o)}$ is the first $n_{o}$
columns of $\bm{I}_o$ (which was defined
immediately after Eq.\ (\ref{eqn:rhf_rdm})).
This can be rearranged as
\begin{align}
      4 \hspace{1mm} ( \bm{h} + \mathcal{A} )
     \hspace{2mm} \bigg[ \hspace{1mm}
   \bm{C}^{(o)}
 \hspace{2mm} \bigg| \hspace{2mm}
   \bm{C}^{(v)}
\bigg] \hspace{2mm}
\bigg[
\begin{array}{c|c}
   \bm{I} - \bm{t} \hspace{0.5mm} \bm{t}^T & \bm{0} \\
     \hline
  \bm{0}  & \bm{t}^T \bm{t} \rule{0pt}{3.8mm}
\end{array}
\bigg]
\label{eqn:term1_intermediate}
\end{align}
Now, use the fact that
\begin{align*}
    \mathcal{A}_{pq} & = 2\mathcal{A}^{coul}_{pq} - \mathcal{A}^{exch}_{pq} \\
    & = C_{rk} C_{sk} \big [2(pq|rs) - (pr|qs)\big] \\
    & = A_{rs} \big [2(pq|rs) - (pr|qs)\big] \\
    & = W[A]_{pq}
\end{align*}
to rewrite Eq.\ (\ref{eqn:term1_intermediate}) as 
\begin{align*}
    4 \hspace{0.5mm}
    \big( \hspace{0.5mm} \bm{h} + \bm{W}[\bm{A}]
    \hspace{0.5mm} \big)
    \hspace{0.5mm} \bm{C}
    \hspace{0.5mm} \gamma^{(MO)}
\end{align*}
which is just Term 1 multiplied by 4.
This factor of 4 cancels with the factor of
4 that appears in front of the
$\bm{S}\bm{C}\bm{\epsilon}$ term when
differentiating the Lagrange multiplier
term from Eq.\ (\ref{eqn:esmf_lagrangian}),
thus delivering Term 1 of
Eq.\ (\ref{eqn:esmf_roothaan}).
Term 2 is gotten similarly by collecting the terms
that involve $\mathcal{B}^{coul}$, $\mathcal{B}^{exch}$,
$\mathcal{C}^{coul}$, or $\mathcal{C}^{exch}$.
Defining $\mathcal{B}$ and $\mathcal{C}$
analogous to $\mathcal{A}$, we get:
\begin{align}
    4 (\mathcal{B} - \mathcal{C}) \bigg[ \hspace{1mm}
    \hspace{1mm}  \bm{C}^{(o)}
 \hspace{2mm} \bigg| \hspace{2mm}
    \bm{0}
 \hspace{1mm} \bigg]  
 \label{eqn:term2_intermediate}
\end{align}
%%%
However, recall that 
\begin{align*}
    (\mathcal{B} - \mathcal{C})_{pq} &= 2\mathcal{B}^{coul}_{pq} - \mathcal{B}^{exch}_{pq} -  2\mathcal{C}^{coul}_{pq} + \mathcal{C}^{exch}_{pq} \\
    & = C_{sb} t_{ib} t_{ia} C_{ra}\big [2 (pq|rs) - (pr|qs)\big] \\& \hspace{4mm}- C_{ri} t_{ia}t_{ja}C_{sj} \big [2(pq|rs) - (pr|qs)\big] \\
    &= \hspace{2mm} \bigg( \bigg[ \hspace{1mm}
   \bm{C}^{(o)}
  \bigg|
   \bm{C}^{(v)}
\bigg] 
\bigg[
\begin{array}{c|c}
  -\bm{t} \hspace{0.5mm} \bm{t}^T & \bm{0} \\
     \hline
  \bm{0}  & \bm{t}^T \bm{t} \rule{0pt}{3.8mm}
\end{array}
\bigg]\bigg[ \hspace{1mm}
   \bm{C}^{(o)}
  \bigg|
   \bm{C}^{(v)}
\bigg] ^T\bigg)_{rs}\\&\hspace{10mm} \cdot \big [2(pq|rs) - (pr|qs)\big]\\
& = W[D]_{pq}
\end{align*}
Then, Eq. (\ref{eqn:term2_intermediate}) can be written as  \begin{align*}
     4 (\mathcal{B} - \mathcal{C}) \bigg[ \hspace{1mm}
    \hspace{1mm}  \bm{C}^{(o)}
 \hspace{2mm} \bigg| \hspace{2mm}
    \bm{0}
 \hspace{1mm} \bigg] = 4 W[D] C A^{(MO)}
\end{align*}
which, again cancelling the factor of 4,
is Term 2. 
Terms 3 and 4 work similarly, this time
collecting the terms involving
$\mathcal{D}^{coul}$ and $\mathcal{D}^{exch}$.

Note that Eq.\ (\ref{eqn:esmf_energy}) can be obtained
by rearranging Eq.\ (\ref{eqn:sqa_energy}).
This follows from regrouping terms the same way as
during the derivation of Eq.\ (\ref{eqn:esmf_roothaan}).

The commutator relation given in
Eq.\ (\ref{eqn:esmf_commutator}) can be derived
from Eq.\ (\ref{eqn:esmf_roothaan}),
analogous to how Eq.\ (\ref{eqn:rhf_commutator})
is derived from Eq.\ (\ref{eqn:rhf_roothaan}).
First, multiply the left hand side of
Eq.\ (\ref{eqn:esmf_roothaan}) by $C^T$ on the left.
Then, transpose the left hand side of
Eq.\ (\ref{eqn:esmf_roothaan}) and 
multiply $C$ on the right.
Finally, take the difference. % of the two expressions.
\begin{align}
  C^T ~ \big(\text{Eq.} (\ref{eqn:esmf_roothaan})~\mathrm{LHS}\big)
- \big(\text{Eq.} (\ref{eqn:esmf_roothaan})~\mathrm{LHS}\big)^T ~ C
\label{eqn:LHS_subtract}
\end{align}
Because $\epsilon$ is symmetric and $\bm{C}^T \bm{S} \bm{C} = I$,
the equivalent manipulations of the right hand sides
of Eq.\ (\ref{eqn:esmf_roothaan}) subtract to give zero.
%At a stationary point of the Lagrangian
%(Eq.\ (\ref{eqn:esmf_lagrangian})),
%the orbital variations are unitary,
%making the multiplier term 0. 
Thus, Eq.\ (\ref{eqn:LHS_subtract}) must be equal to zero,
giving us
\begin{align}
 C^T F_A C \gamma^{(MO)} &- \hspace{1mm} (\gamma^{(MO)})^T C^T F_A^T C
 \notag \\
 + \hspace{1mm} C^T W[D] C A^{(MO)} &- \hspace{1mm} (A^{(MO)})^T C^T (W[D])^T C
 \notag \\
 + \hspace{1mm} C^T W[T]C (T^{(MO)})^T &- \hspace{1mm}T^{(MO)} C^T (W[T])^T C
 \notag \\
 + \hspace{1mm} C^T (W[T])^T C T^{(MO)} &- \hspace{1mm} (T^{(MO)})^T C^T W[T] C
 = 0.
\end{align}
 Recall that $F_A$ is the regular RHF Fock matrix, $W[D]$ is a mean-field operator defined on a difference of density matrices, and $\gamma^{(MO)}$, $A^{(MO)}$ are 1-body RDMs, making these matrices symmetric. For terms with $W[T]$ and $T$, the transposes are part of the expression already. Therefore, this simplifies to the commutator relation as given in Eq. (\ref{eqn:esmf_commutator}). 
 
\begin{table}[b]
\caption{ESMF details in PYCM charge transfer state.\label{tab:pycm_detail}}
\begin{tabular}{c c c c}
\hline
\hline
    Iteration Type   & Energy (a.u.)         &  $\Delta$ (a.u.)         &  DIIS?  \\
\hline
      ORB OPT        &  -571.178433339545    &     3.5554e-01           &   NO    \\
      ORB OPT        &  -571.246925619201    &     2.0214e-01           &   NO    \\
      ORB OPT        &  -571.262163759189    &     1.6952e-01           &   YES   \\
      ORB OPT        &  -571.277606357895    &     2.7521e-02           &   YES   \\
      ORB OPT        &  -571.278746939584    &     9.0729e-03           &   YES   \\
      ORB OPT        &  -571.278998781297    &     3.9852e-03           &   YES   \\
      ORB OPT        &  -571.279081935162    &     2.5653e-03           &   YES   \\
      ORB OPT        &  -571.279097212426    &     8.7319e-04           &   YES   \\
      ORB OPT        &  -571.279100065342    &     4.2506e-04           &   YES   \\
      ORB OPT        &  -571.279100566347    &     1.9735e-04           &   YES   \\
      ORB OPT        &  -571.279100661748    &     8.8807e-05           &   YES   \\
      CIS            &  -571.279215755055    &     N/A                  &   N/A   \\
      ORB OPT        &  -571.279215755013    &     4.9659e-04           &   NO    \\
      ORB OPT        &  -571.279216054833    &     2.4730e-04           &   NO    \\
      ORB OPT        &  -571.279216098757    &     1.8869e-04           &   YES   \\
      ORB OPT        &  -571.279216133376    &     5.1892e-05           &   YES   \\
      CIS            &  -571.279216139390    &     N/A                  &   N/A   \\
\hline
%    $\quad$ & & & & \\
\end{tabular}
\end{table}

Now, starting with Eq.\ (\ref{eqn:esmf_commutator}),
which may not be satisfied for the $\bm{C}$ matrix
we have, we seek to make an improvement by rotating
$\bm{C}$ as $\bm{C}e^{\bm{X}}$,
where $\bm{X}$ is an anti-symmetric matrix.
Assuming that the rotation will be small,
we approximate $e^{\bm{X}} = \bm{I} + \bm{X}$.
%Replacing $C \rightarrow C(I + X)$, the commutator relation can be expressed as 
Inserting this approximation into
Eq.\ (\ref{eqn:esmf_commutator}) gives
\begin{align}
\notag
0 = & \hspace{0.3mm} \big[ \hspace{1mm} (C + CX)^T F_A (C + CX),
      \hspace{1mm} \mathcal{\gamma}^{(MO)} \hspace{1mm} \big] \\
\notag
& + \big[ \hspace{1mm} (C + CX)^TW[D](C + CX),
      \hspace{1mm} A^{(MO)} \hspace{1mm} \big] \\
\notag
& + \big[ \hspace{1mm} (C + CX)^TW[T](C + CX),
      \hspace{1mm} (T^{(MO)})^T \hspace{1mm} \big] \\
& + \big[ \hspace{1mm} (C + CX)^T(W[T])^T(C + CX),
      \hspace{1mm} T^{(MO)} \hspace{1mm} \big].
\end{align}
Dropping any terms that have
more than one power of $\bm{X}$ in them and
rearranging into the form of a linear equation
for $\bm{X}$, we arrive at the working linear
equation that we solve with GMRES to find a
new $\bm{X}$ and then update $\bm{C}$.
%%%
%\begin{align}
%\notag
%0 = \hphantom{+}& \hspace{0.3mm} \big[ \hspace{1mm} F_A^{(MO)} + F_A^{(MO)}X - X F_A^{(MO)},
%      \hspace{1mm} \mathcal{\gamma}^{(MO)} \hspace{1mm} \big] \\
%\notag
%& + \big[ \hspace{1mm}  W[D]^{(MO)} + W[D]^{(MO)}X - X W[D]^{(MO)},
%      \hspace{1mm} A^{(MO)} \hspace{1mm} \big] \\
%\notag
%& + \big[ \hspace{1mm} W[T]^{(MO)} + W[T]^{(MO)}X - X W[T]^{(MO)},
%      \hspace{1mm} (T^{(MO)})^T \hspace{1mm} \big] \\
%& + \big[ \hspace{1mm} W[T]^T^{(MO)} + W[T]^T^{(MO)}X - X W[T]^T^{(MO)},
%      \hspace{1mm} T^{(MO)} \hspace{1mm} \big].
%\end{align}
%Simplifying this expression, 
%\begin{align}
%\notag
%& - \hspace{0.3mm}\big [F_A^{(MO)}, \gamma^{(MO)}\big] - \hspace{0.3mm}\big[W[D]^{(MO)}, A^{(MO)}\big] \\
%\notag
%&- \hspace{0.3mm}\big[W[T]^{(MO)}, T^{(MO)}^T\big] - \hspace{0.3mm}\big[W[T]^T^{(MO)}, T^{(MO)}\big] 
%\notag
%\\\hspace{0.3mm}= \hphantom{+}& \hspace{0.3mm} \big[ \hspace{1mm}  F_A^{(MO)}X - X F_A^{(MO)},
%      \hspace{1mm} \mathcal{\gamma}^{(MO)} \hspace{1mm} \big] \notag\\
%\notag
%& + \big[ \hspace{1mm}   W[D]^{(MO)}X - X W[D]^{(MO)},
%      \hspace{1mm} A^{(MO)} \hspace{1mm} \big] \\
%\notag
%& + \big[ \hspace{1mm}  W[T]^{(MO)}X - X W[T]^{(MO)},
%      \hspace{1mm} (T^{(MO)})^T \hspace{1mm} \big] \\
%& + \big[ \hspace{1mm}   W[T]^T^{(MO)}X - X W[T]^T^{(MO)},
%      \hspace{1mm} T^{(MO)} \hspace{1mm} \big].
%\end{align}
%
%Expanding, 
\begin{align}
\notag
&- \hspace{1mm} \hspace{0.3mm}\big [F_A^{(MO)}, \gamma^{(MO)}\big] - \hspace{0.3mm}\big[W[D]^{(MO)}, A^{(MO)}\big] \\
\notag
& - \hspace{1mm} \hspace{0.3mm}\big[W[T]^{(MO)}, (T^{(MO)})^T\big] - \hspace{0.3mm}\big[(W[T]^{(MO)})^T, T^{(MO)}\big] 
\notag\\
\notag = \hphantom{+} \hspace{0.3mm} & \hspace{1mm}  \big(F_A^{(MO)}X - X F_A^{(MO)}\big)
      \hspace{1mm} \mathcal{\gamma}^{(MO)} \\
&- \hphantom{+} \hspace{0.3mm}  \hspace{1mm} \mathcal{\gamma}^{(MO)}\big(F_A^{(MO)}X - X F_A^{(MO)}\big) \hspace{1mm}  \notag\\
\notag &+ \hphantom{+} \hspace{0.3mm}  \hspace{1mm} \big(W[D]^{(MO)}X - X W[D]^{(MO)}\big)
      \hspace{1mm} A^{(MO)} \\
&- \hphantom{+} \hspace{0.3mm}  \hspace{1mm} A^{(MO)}\big(W[D]^{(MO)}X - X W[D]^{(MO)}\big) \hspace{1mm} \notag \\
\notag &+  \hphantom{+} \hspace{0.3mm}  \hspace{1mm}  \big(W[T]^{(MO)}X - X W[T]^{(MO)}\big)
      \hspace{1mm} (T^{(MO)})^T \hspace{1mm} \notag \\
&-  \hphantom{+} \hspace{0.3mm}  \hspace{1mm}(T^{(MO)})^T \big(W[T]^{(MO)}X - X W[T]^{(MO)}\big) \notag\\
 &+  \hphantom{+} \hspace{0.3mm}  \hspace{1mm}  \big((W[T]^{(MO)})^TX - X (W[T]^{(MO)})^T\big)
      \hspace{1mm} T^{(MO)} \notag \\
&- \hphantom{+} \hspace{0.3mm}  \hspace{1mm}T^{(MO)}\big((W[T]^{(MO)})^TX - X (W[T]^{(MO)})^T\big)  \hspace{1mm}.
\label{eqn:linear_eq_x}
\end{align}
Note that, once we use $\bm{X}$ to update $\bm{C}$,
we reevaluate the mean-field operators
$F_A$, $W[D]$, and $W[T]$.
%Instead, we keep these AO-basis mean-field operators
%fixed, use the new $\bm{C}$ to rotate them to
%the MO basis, and then solve Eq.\ (\ref{eqn:linear_eq_x})
%again for a new rotation.
%Only after we have done this multiple times
%and made the error in Eq.\ (\ref{eqn:esmf_commutator})
%small (as measured for example by by $\Delta$, see next
%section) do we perform Fock builds to update
%the AO-basis mean-field operators.
Once multiple SCF orbital optimization iterations
are complete and the
error in Eq.\ (\ref{eqn:esmf_commutator}) is
small (as measured for example by by $\Delta$, see next
section),
we perform a new CIS calculation in the new
orbital basis to
update $\bm{t}$, and then start in again on
orbital optimization via our linear equation.
As a final note, we use a diagonal preconditioner
to approximate the inverse of the linear
transformation and accelerate
the GMRES solution of the linear equation.
For elements of $\bm{X}$ in the occupied-occupied
or virtual-virtual blocks, the preconditioner's
diagonal element is just one.
For elements of $\bm{X}$ in the occupied-virtual
or virtual-occupied blocks,
i.e.\ $X_{ia}$ or $X_{ai}$,
the preconditioner's diagonal element is
\begin{align}
\bigg(
  \left[F^{(MO)}_A\right]_{aa}
- \left[F^{(MO)}_A\right]_{ii}
\bigg)^{-1}
\end{align}

%This is the linear equation in $X$ that is then solved via GMRES. 

 %    = 2 A^{coul}_{tq} U_{qb} t^T_{bi} t_{ic} + 2 B^{coul}_{tr} U_{rm}  = \boxed{2A^{coul} U_v t^T t + 2 B^{coul} U_o}

%$\quad$
%
%$\quad$
%
%$\quad$
%
%$\quad$
%
%$\quad$
%
%$\quad$
%
%$\quad$
%
%$\quad$
%
%$\quad$
%
%$\quad$
%
%$\quad$
%
%$\quad$
%
%$\quad$
%
%$\quad$
%
%$\quad$
%
%$\quad$
%
%$\quad$
%
%$\quad$
%
%\newpage

\section{Charge Transfer Iteration Details}
\label{sec:iter_details}

For the charge transfer optimization in PYCM, we show
in Table \ref{tab:pycm_detail} the total energy
as well as $\Delta$, the Frobenius norm
of the error in the commutator matrix equation in the
molecular orbital basis,
at each stage of ESMF's SCF optimization.
Each ``ORB OPT'' iteration corresponds to one
solution of the linear equation for X.
The ``DIIS?'' column labels which
orbital optimization steps employed DIIS.
After optimal orbitals for the current
$\bm{t}$ coefficients are found,
a CIS calculation is performed in the new
molecular orbital basis in order to update
$\bm{t}$.
Note that the initial guess in this case
was to set the $\bm{C}$ and $\bm{t}$
matrices to correspond
to a HOMO$\rightarrow$LUMO transition
in the RHF orbital basis, which corresponds
to the charge transfer state in question.
The orbitals were then optimized by our
SCF procedure, after which a CIS calculation
updated $\bm{t}$, after which the orbitals
were optimized again, after which another
CIS calculation shows that the energy has
converged.

%$\vspace{1mm}$

%$\vspace{1mm}$

%$\vspace{1mm}$

%\section{Molecular Geometries}
%\label{sec:geometries}

%\noindent
%Note: molecular geometries can be found
%below Table \ref{tab:pycm_detail}.

%\newpage

%$\vspace{1mm}$

%$\vspace{1mm}$

\section{Geometries}
\label{sec:geom}

Molecular geometries
(some in Angstrom and some in Bohr, see table labels)
are given in the following tables.
Note that the PYCM geometry is on the next page.

\begin{table}[h!]
\begin{tabular}{l r@{.}l r@{.}l r@{.}l}
\hline
\hline
\multicolumn{7}{l}{Molecule:  Water
                   $\quad$Units: Angstrom} \\
\hline
O & 0&0000000 &  0&0000000 &  0&1157190 \\
H & 0&0000000 &  0&7487850 & -0&4628770 \\
H & 0&0000000 & -0&7487850 & -0&4628770 \\
\hline
\end{tabular}
\end{table}

\begin{table}[h!]
\begin{tabular}{l r@{.}l r@{.}l r@{.}l}
\hline
\hline
\multicolumn{7}{l}{Molecule:  Ethylene
                   $\qquad$Units: Angstrom} \\
\hline
 C & 0&0000000000 &      0&0000000000 &      0&6727698502 \\
 C & 0&0000000000 &      0&0000000000 &     -0&6727698502 \\
 H & 0&0000000000 &     -0&9347680531 &      1&2426974978 \\
 H & 0&0000000000 &      0&9347680530 &      1&2426974980 \\
 H & 0&0000000000 &     -0&9347680532 &     -1&2426974978 \\
 H & 0&0000000000 &      0&9347680530 &     -1&2426974980 \\
\hline
\end{tabular}
\end{table}

\begin{table}[h!]
\begin{tabular}{l r@{.}l r@{.}l r@{.}l}
\hline
\hline
\multicolumn{7}{l}{Molecule:  Formaldehyde
                   $\qquad$Units: Angstrom} \\
\hline
 C & -0&0910041349 &      0&1032665042 &     -0&6906844847 \\
 O & -0&7161280585 &      0&7732151716 &      0&1085248268 \\
 H & -0&2415602237 &      0&1918899655 &     -1&7934092839 \\
 H &  0&6751090789 &     -0&6449677992 &     -0&3749139325 \\
\hline
\end{tabular}
\end{table}

%\begin{table*}
%\begin{tabular}{l r@{.}l r@{.}l r@{.}l}
%\hline
%\hline
%\multicolumn{7}{l}{Molecule:  Aniline
%                   $\qquad$Units: Angstrom} \\
%\hline
% H &  -0&9886445675 &  -1&2053993263 &   2&3426597492 \\
% C &  -0&7254028224 &  -0&7546403472 &   1&3853311333 \\
% C &  -1&6886232268 &  -0&6412745623 &   0&3587705746 \\
% H &  -2&7069947897 &  -0&9973968638 &   0&5229639685 \\
% C &  -1&3530135115 &  -0&0573891758 &  -0&8791629948 \\
% H &  -2&1073651799 &   0&0279829694 &  -1&6659481278 \\
% C &  -0&0338546677 &   0&4075876983 &  -1&1197229594 \\
% N &   0&2951387202 &   1&0269124655 &  -2&3414318389 \\
% H &  -0&3259490156 &   0&9001172199 &  -3&1287536559 \\
% H &   1&2715708517 &   1&1518346857 &  -2&5709485881 \\
% C &   0&9290414462 &   0&3019608377 &  -0&0824556606 \\
% H &   1&9460313495 &   0&6663189894 &  -0&2508036518 \\
% C &   0&5853344550 &  -0&2831435357 &   1&1526789889 \\
% H &   1&3410018586 &  -0&3599172546 &   1&9362196629 \\
%\hline
%\end{tabular}
%\end{table*}

\begin{table}[h!]
\begin{tabular}{l r@{.}l r@{.}l r@{.}l}
\hline
\hline
\multicolumn{7}{l}{Molecule:  Toluene
                   $\qquad$Units: Angstrom} \\
\hline
 C &  0&0865862990 &  0&7410453762 &  0&0688914923 \\
 C & -0&1345668535 & -0&6531059507 &  0&0602786905 \\
 C & -1&4505649626 & -1&1603576134 & -0&0093490679 \\
 C & -2&5405851883 & -0&2644802090 & -0&0661246831 \\
 C & -2&3124601785 &  1&1285606766 & -0&0571533440 \\
 C & -0&9977830949 &  1&6502417205 &  0&0134992344 \\
 C & -0&7535284502 &  3&1540596435 & -0&0111519475 \\
 H &  1&1079021258 &  1&1261067270 &  0&1292876906 \\
 H &  0&7134665908 & -1&3382132205 &  0&1093728295 \\
 H & -1&6243218250 & -2&2374292057 & -0&0134283434 \\
 H & -3&5611922298 & -0&6477509085 & -0&1151906890 \\
 H & -3&1617587236 &  1&8157708718 & -0&0950213452 \\
 H & -0&6369699217 &  3&5271815420 & -1&0399480108 \\
 H &  0&1595927745 &  3&4107593481 &  0&5443343618 \\
 H & -1&5935598619 &  3&6938717021 &  0&4484031317 \\
\hline
\end{tabular}
\end{table}

\begin{table}[h!]
\begin{tabular}{l r@{.}l r@{.}l r@{.}l}
\hline
\hline
\multicolumn{3}{l}{Molecule:  PYCM} &
\multicolumn{4}{l}{Units: Bohr} \\
\hline
C &  7&48914118998055 &  -2&05146035486422 &  -1&28980829449057 \\ 
C &  5&98770398368311 &   0&04915735478248 &  -0&06865458045111 \\
C &  7&61081607331134 &   2&27669871460533 &   0&67560530219819 \\
C &  3&46706532904223 &  -0&09116628967473 &   0&30637708021880 \\
C &  1&92647158203893 &  -2&41958847844936 &  -0&43805415580348 \\
C & -0&75455162629890 &  -2&43014371114785 &   0&69884575028668 \\
C & -2&18486749739264 &  -0&06198337612267 &   0&06973201498199 \\
C & -0&64398578220618 &   2&33392846576440 &   0&05840028021288 \\
C &  1&88958084771184 &   2&01941654524793 &   1&44673369015458 \\
C & -4&70904737992718 &  -0&06309418170513 &  -0&46180929064412 \\
C & -6&17576878134698 &  -2&34798184267961 &  -0&48999098912292 \\
N & -7&35364877360203 &  -4&20620382240250 &  -0&50945688691185 \\
C & -6&03132613865057 &   2&24217192107952 &  -1&01894902061744 \\
N & -7&06517180189986 &   4&13250579177077 &  -1&46439337259584 \\
H &  8&54937612247998 &  -1&31872561024836 &  -2&93252209992581 \\
H &  6&33486664187961 &  -3&64092303587743 &  -1&95285217876895 \\
H &  8&91885629767513 &  -2&80382984977844 &   0&03384049526409 \\
H &  9&17826984585370 &   1&64141995115378 &   1&89968164549696 \\
H &  6&59508025455693 &   3&76874545170370 &   1&69452668746078 \\
H &  8&50253869488760 &   3&13726625417923 &  -1&00559579441233 \\
H &  2&88064720049037 &  -4&16472134045315 &   0&16752333485011 \\
H &  1&76036799956004 &  -2&54566120290402 &  -2&51696904366760 \\
H & -0&60805363262071 &  -2&53086096937472 &   2&78367809044349 \\
H & -1&80585658070689 &  -4&12167616649262 &   0&11092207863213 \\
H & -1&74995920022716 &   3&90829005165278 &   0&85174533709869 \\
H & -0&27696758321317 &   2&85096221165140 &  -1&93484378989024 \\
H &  1&49786269333475 &   1&62230247974489 &   3&46233841913685 \\
H &  2&89384261982421 &   3&83106450894856 &   1&40264023881446 \\
\hline
\end{tabular}
\end{table}

\end{document}